\documentclass[submitting]{nst}
\usepackage{subfigure,dcolumn}
\usepackage[T2A,T1]{fontenc}
\usepackage{listings}
\usepackage{longtable}
\usepackage{grffile} 
\usepackage{graphics}
\usepackage{graphicx} 
\usepackage{amsmath}    
\usepackage{hyperref}   
\usepackage{epsfig}
\usepackage{epstopdf}
\usepackage{color}
\usepackage{url}
\usepackage{bm}

\newcommand{\ber}{\begin{eqnarray}}
\newcommand{\eer}{\end{eqnarray}}
\newcommand{\pt}{p_{\rm T}} 
\newcommand{\snn}{\sqrt {s_{\rm NN}}}
\newcommand{\tauf} {\tau_{_{\rm F}}}
\newcommand{\tone}{t_1}
\newcommand{\ttwo}{t_2} 
\newcommand{\detr} {dE_{_{\rm T}}}
\newcommand{\sigs}{\sigma_{\rm soft}}
\newcommand{\vtwoep}{v_2\{\rm EP\}}
\newcommand{\dtwoetr} {d^2E_{_{\rm T}}}

\begin{document}

\title{Further developments of a multi-phase transport model for relativistic nuclear collisions}
\thanks{Z.-W.L. is supported in part by the National Science 
  Foundation under Grant No. PHY-2012947. L.Z. is supported in part by the
  National Natural Science Foundation of China under Grant
  No. 11905188.}

\author{Zi-Wei Lin}
\email[Corresponding author, ]{linz@ecu.edu}
\affiliation{Department of Physics, East Carolina University, Greenville, North Carolina 27858, USA} 
\author{Liang Zheng}
\affiliation{School of Mathematics and Physics, China University of Geosciences (Wuhan), Wuhan 430074, China}

\begin{abstract}
A multi-phase transport (AMPT) model was constructed as a
self-contained kinetic theory-based description of 
relativistic nuclear collisions as it contains four main
components: the fluctuating initial condition, a parton cascade, 
hadronization, and a hadron cascade. 
Here, we review the main developments after the first public release of the
AMPT source code in 2004 and the corresponding publication that
described the physics details of the model at that time. 
We also discuss possible directions for future developments of
the AMPT model to better study the properties of the dense matter
created in relativistic collisions of small or large systems.
\end{abstract}


\maketitle

\begin{nolinenumbers}

\section{Introduction}

In high energy heavy ion collisions~\cite{Lee:1974ma}, a hot and dense
matter made of parton degrees of freedom, the quark-gluon plasma
(QGP), has been expected to be created~\cite{Shuryak:1978ij}. 
Experimental data from the Relativistic Heavy Ion Collider (RHIC) 
and the Large Hadron Collider (LHC) 
\cite{Arsene:2004fa,Back:2004je,Adams:2005dq,Adcox:2004mh,Heinz:2013th,Busza:2018rrf} 
strongly indicate that the QGP is indeed created in heavy ion
collisions at high energies~\cite{Gyulassy:2004zy}. 
Comprehensive comparisons beween the experimental data and theoretical
models are essential for the extraction of key properties of the high
density matter, including the structure of the QCD 
phase diagram at high temperature and/or high net-baryon density.  
Many theoretical models including transport
models~\cite{Bass:1998ca,Zhang:1999bd,Xu:2004mz,Lin:2004en,Cassing:2009vt},
hydrodynamic
models~\cite{Huovinen:2001cy,Betz:2008ka,Schenke:2010rr,Bozek:2011if},
and hybrid models~\cite{Petersen:2008dd,Werner:2010aa,Song:2010mg}
have been constructed to simulate and study the phase space evolution
of the QGP.

A multi-phase transport (AMPT) model \cite{Lin:2004en} is one such model. 
The AMPT model aims to apply the kinetic theory approach to 
describe the evolution of heavy-ion collisions as it
contains four main components: the fluctuating initial condition,
partonic interactions, hadronization, and hadronic interactions.  
The default version of the AMPT model \cite{Zhang:1999bd,Lin:2000cx}
was first constructed. Its initial condition is based on the 
Heavy Ion Jet INteraction Generator (HIJING) two-component
model~\cite{Wang:1991hta,Gyulassy:1994ew}, 
then minijet partons enter the parton cascade and eventually recombine
with their parent strings to hadronize via the Lund string 
fragmentation~\cite{Sjostrand:1993yb}. The default AMPT model can well
describe the rapidity distributions and transverse momentum ($\pt$)
spectra of  identified particles observed in heavy ion collisions at
SPS and RHIC. However, it significantly underestimates the elliptic
flow ($v_2$) at RHIC. 

Since the matter created in the early stage of high energy heavy ion
collisions is expected to have a very high energy density and thus 
should be in parton degrees of freedom, the string
melting version of the AMPT (AMPT-SM) model \cite{Lin:2001zk} 
was then constructed, where all the excited strings from a heavy ion
collision are converted into partons and a spatial quark coalescence
model is invented to describe the hadronization process. 
String melting increases the parton density and produces an
over-populated partonic matter~\cite{Lin:2014tya}, while 
quark coalescence further enhances the elliptic flow of
hadrons~\cite{Lin:2001zk,Molnar:2003ff}.  
As a result, the string melting AMPT model is able to describe the
large elliptic flow in Au+Au collisions at RHIC energies with a
rather small parton cross section~\cite{Lin:2001zk,Molnar:2019yam}.  

The source code of the AMPT model was first publicly released online  
around April 2004, and a subsequent publication \cite{Lin:2004en}  
provided detailed descriptions of the model such as the included physics 
processes and modeling assumptions.  
The AMPT model has since been widely used to simulate the evolution of the 
dense matter created in high energy nuclear collisions. 
In particular, the string melting version of the AMPT
model~\cite{Lin:2001zk,Lin:2004en} 
can well describe the anisotropic flows and particle correlations 
in collisions of small or large systems at both RHIC and LHC
energies~\cite{Lin:2001zk,Lin:2004en,Bzdak:2014dia,Ma:2016fve,He:2017tla,Zhang:2018ucx}. 
The AMPT model is also a useful test bed of different ideas. For
example, the connection between the triangular flow and initial
geometrical fluctuations was discovered with the help of AMPT
simulations~\cite{Alver:2010gr}, 
and the model has also been applied to studies of vorticity and
polarization in heavy ion
collisions~\cite{Jiang:2016woz,Li:2017slc,Lan:2017nye}.
 
Experimental data from heavy ion collisions fit with
hydrodynamics-inspired models suggest that particles are locally
thermalized and possess a common radial flow
velocity~\cite{Abelev:2008ab}.  Large momentum anisotropies such as
the elliptic flow~\cite{Ollitrault:1992bk} have been measured 
in large collision systems, as large as the hydrodynamics
predictions~\cite{Heinz:2013th,Gale:2013da}.  This suggests that the
collision system is strongly interacting and close to 
local thermal equilibrium~\cite{Gyulassy:2004zy}. 
Transport models can also generate large anisotropic flows. 
The string melting AMPT model~\cite{Lin:2001zk,Lin:2004en} can
describe the large anisotropic flows with a rather small parton cross
section of $\sim 3$ mb ~\cite{Lin:2001zk} and the flow enhancement from
quark
coalescence~\cite{Lin:2001zk,Molnar:2003ff,Li:2016ubw,Li:2016flp,Molnar:2019yam}.  
Without the quark coalescence, a pure parton transport for
minijet gluons requires an unusually large parton cross section of
$\sim 40-50$ mb~\cite{Molnar:2001ux,Molnar:2019yam} 
for the freezeout gluons to have a similar magnitude of elliptic flow
as charged hadrons in the experiments. 
This minijet gluon system, despite a factor of $\sim 2.5$ lower parton
multiplicity at mid-rapidity, has a factor of $\sim 6$
smaller mean free path than the string melting AMPT model for $200A$
GeV Au+Au collisions at impact parameter $b=8$
fm~\cite{Molnar:2019yam}. In general, for large systems at high
energies transport models tend to approach hydrodynamics since the
average number of collisions per particle is large  and thus the bulk
matter is close to local equilibrium. Hydrodynamics models and
transport models are also complementary to each other. For example,
hydrodynamics models provide a direct access  to the equation of state
and transport coefficients, while transport models can address
non-equilibrium dynamics and provide a microscopic picture of the
interactions. 

Recent data from small systems, however, hint at significant
anisotropic flows in high multiplicity $pp$ and $p$Pb collisions at
the LHC~\cite{Khachatryan:2010gv} and $p/d/^3$He+Au collisions at
RHIC~\cite{Adare:2014keg,PHENIX:2018lia}.  
Hydrodynamic calculations seem to describe the experimental data
well~\cite{Bozek:2010pb,Bozek:2012gr}. The AMPT-SM model also seems to
describe the measured correlations~\cite{Bzdak:2014dia}.  
This suggests that the collision of these small systems might create a
QGP as well, in contrast to naive expectations. 
On the other hand, it is natural to expect hydrodynamic models and
transport models to be different for small colliding systems due
to non-equilibrium  dynamics.  Indeed, recently it has been realized
that parton transport can convert the initial spatial anisotropies
into significant anisotropic flows in the momentum space through the
parton escape mechanism~\cite{He:2015hfa,Lin:2015ucn}, especially in
small systems where the average number of collisions per particle is
small. Kinetic theory studies also show that a single 
scattering is very efficient in changing the particle momentum 
distribution~\cite{Kurkela:2018ygx}.   There are also many studies on
whether and how hydrodynamics could be 
applicable to small systems~\cite{Heinz:2019dbd,Schenke:2021mxx}.  
In addition, there are active debates on whether the momentum
anisotropies in small systems mainly come from initial state 
correlations~\cite{Dusling:2017dqg,Mace:2018vwq} or final state
interactions~\cite{He:2015hfa,Lin:2015ucn,Weller:2017tsr,Kurkela:2018ygx,Kurkela:2019kip}. 
Furthermore, the differences between the anisotropic flow data of
small systems from different collaborations still need to be fully
resolved~\cite{PHENIX:2018lia,Lacey:2020ime,PHENIX:2021bxz}. 
Therefore, the system size dependence of various observables,
particularly the anisotropic flows from small to large systems, 
could provide key information on the origin of collectivity.

The paper is organized as follows. After the introduction, 
we review in Sect.~\ref{sec:develop} the main developments of the
AMPT model after the first public release of its source code in 2004
~\cite{oscar,ampt,Lin:2004en}.
They include the addition of deuteron productions in
Sect.~\ref{subsec:deu}, the string melting model that can
simultaneously reproduce the yield, transverse momentum spectra and
elliptic flow of the bulk matter in heavy ion collisions in
Sect.~\ref{subsec:sm}, the new quark coalescence model in
Sect.~\ref{subsec:coal}, incorporation of the finite nuclear thickness
along beam directions in Sect.~\ref{subsec:width}, 
incorporation of modern parton distribution functions of nuclei
in Sect.~\ref{subsec:pdf}, improved treatment of heavy quark  
productions in Sect.~\ref{subsec:hf}, the introduction of local nuclear
scaling of key input parameters to describe the system size dependence
in Sect.~\ref{subsec:local}, incorporation of PYTHIA8 and nucleon 
substructure in the initial condition in Sect.~\ref{subsec:pythia8},
and benchmark and improvement of the parton cascade algorithm in 
Sect.~\ref{subsec:zpc}.
We then briefly review other developments of the
AMPT model in Sect.~\ref{sec:other}. 
Finally, in Sect.~\ref{sec:summary}, we summarize and discuss possible 
directions for further developments of the AMPT model.

\section{Main developments}
\label{sec:develop}

We now review the main developments of the AMPT model 
after the first public release of the AMPT source code in 2004
~\cite{oscar,ampt} and the corresponding publication that
described the physics details of the model at that
time~\cite{Lin:2004en}. 
These developments are listed mostly in chronological 
order. In terms of the four main components of the
AMPT model, Sects.~\ref{subsec:sm}, \ref{subsec:width}, 
\ref{subsec:pdf}, \ref{subsec:hf}, \ref{subsec:local},
\ref{subsec:pythia8} are about the initial condition, 
Sect.~\ref{subsec:zpc} is about the parton cascade, 
Sect.~\ref{subsec:coal} is about the hadronization, 
while Sect.~\ref{subsec:deu} is about the hadron cascade. 
Currently the public versions of the AMPT model
since v1.26t5/v2.26t5~\cite{ampt} have incorporated the changes made
in the developments described in Sects. \ref{subsec:deu} and
\ref{subsec:sm}; changes from the other developments will be
released in the future.

\subsection{Deuteron productions in the hadron cascade}
\label{subsec:deu}

Light nuclei such as deuteron ($d$) and triton ($t$) are produced and
observed in high energy nuclear collisions at RHIC and
LHC~\cite{Abelev:2010rv,Sharma:2011ya}.   
They have been proposed to be important for the search of the QCD
critical
point~\cite{Sun:2017xrx,Shuryak:2018lgd,Bzdak:2019pkr,Luo:2020pef} 
and thus the study of light nuclei has become more active recently. 
Currently the production mechanism of light nuclei is still under
debate, as there are several different models that describe the data
including the statistical
model~\cite{Cleymans:1992zc,Braun-Munzinger:2015hba}, the nucleon
coalescence 
model~\cite{Sato:1981ez,Csernai:1986qf,Gutbrod:1988gt,Zhu:2015voa,Sun:2017ooe},
and dynamical models based on the kinetic
theory~\cite{Danielewicz:1991dh,Oh:2007vf,Oh:2009gx}.

\begin{figure}[!htb]
\includegraphics [width=0.9\hsize]  {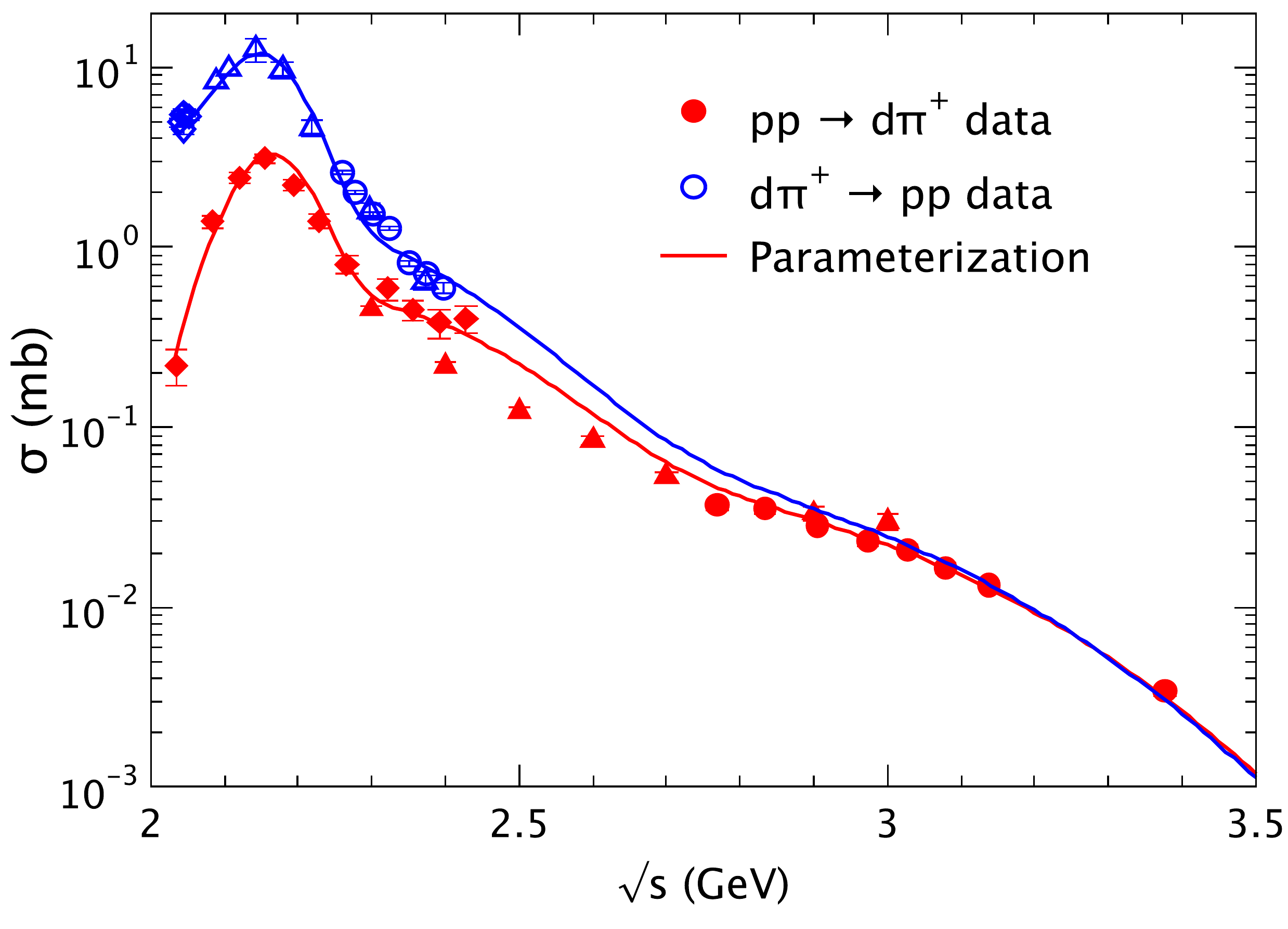}
\caption{Experimental data on the total cross sections of $p p \to d
\pi^+$~\cite{Heinz:1968zz,Anderson:1974tp,Shimizu:1982dx} (filled
symbols) and $d \pi^+ \to p
p$~\cite{Boswell:1982xm,Borkovsky:1985fb,Gogolev:1993ay}  (open
symbols) in comparison with our parameterizations (solid curves).} 
\label{fig:sigmad}
\end{figure}

We have modified the AMPT model to provide a kinetic theory
description of deuteron production and annihilation 
by adding the following reactions~\cite{Oh:2009gx}:
\begin{equation}
B B' \leftrightarrow M d,
\label{eq:md}
\end{equation}
where $M = \pi$, $\rho$, $\omega$, and $\eta$, while $B$ and $B'$
stand for baryons $N$, $\Delta$, $P_{11}(1440)$, and $S_{11}(1535)$. 
Note that the hadron cascade component of the AMPT
model~\cite{Lin:2004en}, based on a relativistic transport (ART) model
~\cite{Li:1995pra,Li:1996rua,Li:2001xh}, already includes the
interactions of $\pi$, $K$, $\eta$, $\rho$, $\omega$, $\phi$, $K^*$, $N$,
$\Delta(1232)$, $P_{11}(1440)$, $S_{11}(1535)$ as well as their
antiparticles. 
For the cross sections of the reactions $BB'\to Md$, we assume that
their angular integrated mean squared matrix elements that are
averaged over initial and summed over final spins and isospins are the
same as that for the reaction $N N \to d \pi$ at the same center of
mass energy $\sqrt{s}$. The cross sections for the inverse reactions
$Md\to BB'$ are then determined from the detailed balance. 
In addition to the production and annihilation processes for
deuterons, we also include their elastic scatterings with mesons $M$
and baryons $B$~\cite{Oh:2009gx}. 

Experimentally, the cross sections for both the reaction $p p \to d
\pi^+$~\cite{Heinz:1968zz,Anderson:1974tp,Shimizu:1982dx} 
and the reaction $\pi^+ d \to p
p$~\cite{Boswell:1982xm,Borkovsky:1985fb,Arndt:1993nd,Gogolev:1993ay} have 
been extensively measured, and the former is  given by 
\begin{eqnarray}
\sigma(pp\to d\pi^+) = \frac14 \frac{p_\pi^{}}{p_N^{}} f(s),
\label{eq:pp}
\end{eqnarray}
where $p_N^{}$ and $p_\pi^{}$ are, respectively, the magnitude of the three-momenta
of initial and final particles in the center of mass frame. The function $f(s)$, which is proportional to the angular integrated mean squared matrix elements that are summed over initial and final spins for the reaction $p p \to \pi^+ d$, is parameterized as 
\begin{eqnarray}
f(s) &=& 26 \exp[-(s-4.65)^2/0.1] + 4 \exp[-(s-4.65)^2/2]
\nonumber \\ 
&&+ 0.28\exp[-(s-6)^2/10],
\end{eqnarray}
where $\sqrt{s}$ is in the unit of GeV and
$f(s)$ is in the unit of mb. For the inverse reaction $d \pi^+ \to
pp$, its cross section is related to that for $pp\to d\pi^+$ via the
detailed balance relation:
\begin{equation}
\sigma(d \pi^+ \to pp) = \frac {2 p_N^2}{3p_\pi^2} \sigma (pp\to d\pi^+).
\end{equation}
These parameterizations are compared
with the experimental data in Fig.~\ref{fig:sigmad}. The cross
sections for the isospin averaged reactions $NN\to d\pi$ and $\pi d\to
NN$ can then be obtained as $\sigma(NN\to d\pi)=3\sigma(pp\to
d\pi^+)/4$ and $\sigma(d\pi\to NN)=\sigma(d\pi^+\to pp)$.

We have coupled the above deuteron transport 
with an initial hadron distribution after hadronization 
as parameterized by a blast wave model~\cite{Oh:2009gx}, 
where a nucleon coalescence model using the deuteron Wigner function
~\cite{Gyulassy:1982pe} was also applied for comparison. 
We find that the transport model gives very similar deuteron $\pt$
spectra as the coalescence model; however the elliptic flows from the
two models are different. 
In particular, the transport model gives a deviation of the elliptic
flow from the exact nucleon number scaling 
at relatively high $\pt$ and agrees better with the measured data.

On the other hand, the deuteron yield obtained directly from the
AMPT-SM model is typically much lower than the experimental data. This
could be due  to the assumed relation between the $B B'
\leftrightarrow M d$ and $p p \leftrightarrow d \pi$ cross 
sections, which can be further constrained by using the measured total
$\pi d$ cross section, or the lack of additional production channels
such as  $\pi n p \leftrightarrow  \pi d$~\cite{Oliinychenko:2018ugs}. 
The low yield could also be partly due to the assumption of no
primordial deuteron formation from quark coalescence. 
There are also studies~\cite{Zhu:2015voa,Sun:2020uoj}
that applied the nucleon coalescence model to the kinetic freezeout
nucleon distributions from the AMPT-SM model. 
It has been found that the resultant light nuclei yields depend sensitively on
the freezeout surface, which is affected by both the partonic expansion
and the hadronization (quark coalescence) criterion. The yields also
depend on the coalescence function used for the light
nuclei~\cite{Sun:2020uoj}, especially for small collision systems
where the suppression due to the light nuclei size 
~\cite{Sun:2018mqq} could be significant.
Further improvements of the AMPT model regarding the deuteron cross
sections, the parton phase, and the hadronization criterion will
benefit the studies of light nuclei.
 
\subsection{String melting model to describe the bulk matter}
\label{subsec:sm}

\begin{figure*}[!htb]
\centering
\includegraphics [width=0.9\hsize]  {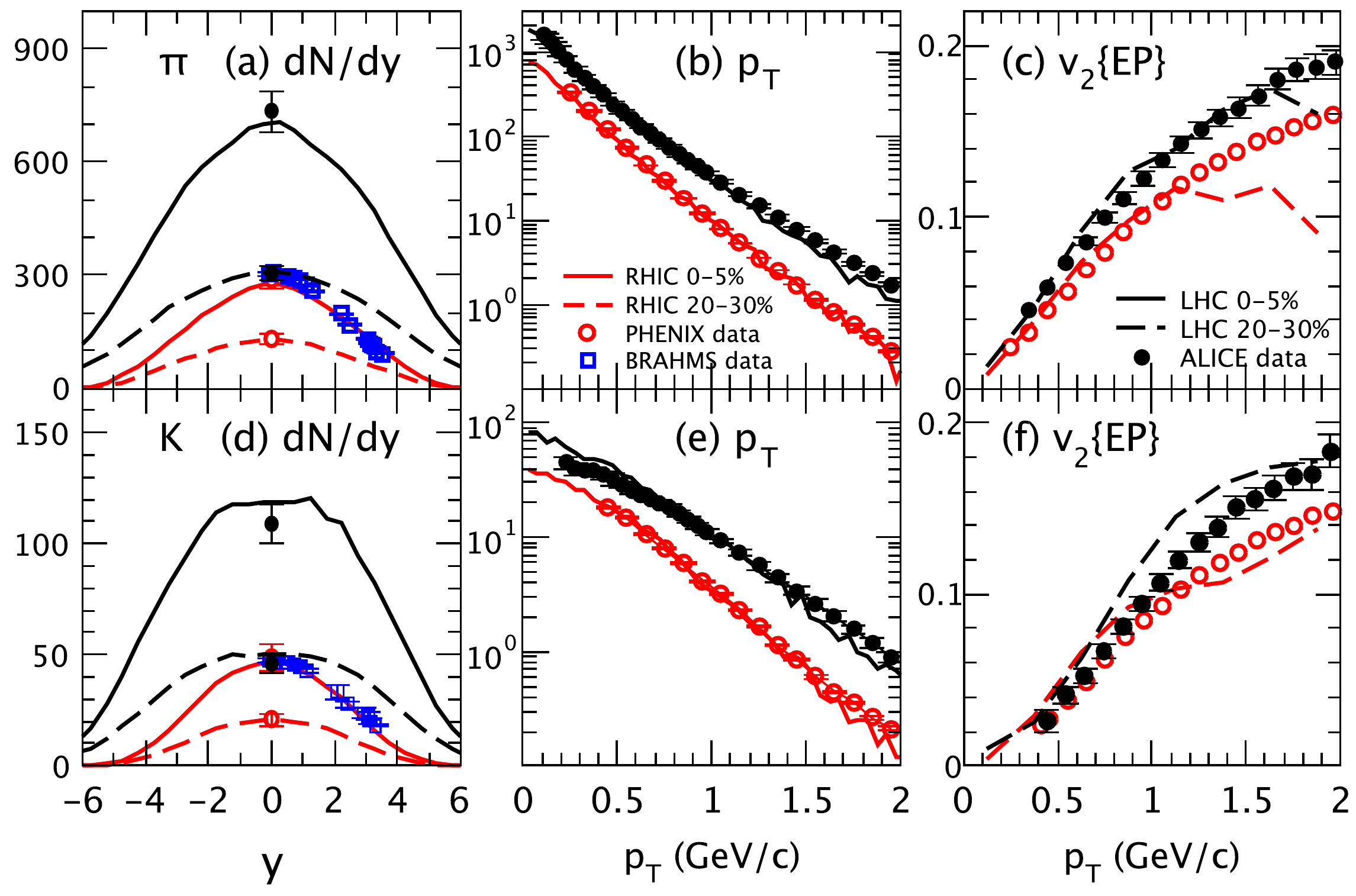}
\caption{AMPT-SM results for pions (upper panels) and kaons (lower
  panels) on dN/dy of (a) $\pi^+$ and (d) $K^+$ in central and mid-central
collisions, $\pt$ spectra $dN/(2\pi \pt d\pt dy)$ in the unit of
$c^2/{\rm GeV}^2$ of (b) $\pi^+$ and (e) $K^+$ at mid-rapidity in central collisions, 
and elliptic flow $\vtwoep$ of (c) charged pions and (f) charged kaons at mid-rapidity 
in mid-central collisions in comparison with the experimental data for
central (0-5\%) and/or mid-central (20-30\%) Au+Au collisions at
$200A$ GeV and Pb+Pb collisions at $2.76A$ TeV. 
}
\label{fig:sm}
\end{figure*}

The Lund string model \cite{Sjostrand:1993yb} is used 
in both the default and string melting versions of the AMPT model. 
In the default AMPT model, minijet partons recombine with their parent
strings after the parton cascade to hadronize via the Lund string
model into primordial hadrons. In the AMPT-SM model, the
primordial hadrons that would be produced from the excited Lund
strings in the HIJING model are ``melt'' into primordial quarks and
antiquarks. Therefore, the parameters in the Lund string model affect
the AMPT model results. 
In the Lund model, one assumes that a string
fragments into quark-antiquark pairs with a Gaussian distribution in
transverse momentum. 
Hadrons are formed from these quarks and antiquarks, with the 
longitudinal momentum given by the Lund symmetric fragmentation
function~\cite{Andersson:1983jt,Andersson:1983ia}
\begin{equation}
f(z) \propto z^{-1}(1-z)^{a_L}~e^{-b_L m^2_{\rm T}/z}.
\label{eq:lund} 
\end{equation} 
In the above, $z$ represents the light-cone momentum fraction of the
produced hadron with respect to that of the fragmenting string and
$m_{\rm T}$ is the transverse mass of the hadron. 

When using the HIJING values~\cite{Wang:1991hta,Gyulassy:1994ew} for
the key Lund string fragmentation parameters, $a_L=0.5$ and $b_L=0.9$
GeV$^{-2}$,  the default AMPT model works well for particle yields
and $\pt$ spectra in $pp$ collisions at various energies.  
However, it gives too small a charged particle yield in 
central Pb+Pb collisions at the SPS energy of $E_{\rm
  LAB}=158A$ GeV~\cite{Zhang:1999bd,Lin:2000cx}. 
Instead, modified values of $a_L=2.2$ and $b_L=0.5$
GeV$^{-2}$ were needed to fit the charged particle yield and $\pt$
spectra in Pb+Pb collisions at SPS.  
For heavy ion collisions at higher energies such as RHIC energies, 
the default version of the AMPT model with these parameter values also 
reasonably describes hadron dN/d$\eta$, dN/dy and the $\pt$ spectra in 
heavy ion collisions, although it underestimates the elliptic
flow~\cite{Lin:2001zk}. 

On the other hand, the AMPT-SM model~\cite{Lin:2001zk,Lin:2004en}, due
to its dense parton phase and quark coalescence, reasonably describes
the elliptic flow~\cite{Lin:2001zk} 
and two-pion interferometry~\cite{Lin:2002gc} in heavy ion
collisions. However, the versions before 2015~\cite{ampt} (i.e.,
before v2.26t5)  could not reproduce well the hadron dN/d$\eta$, dN/dy
and $\pt$ spectra (when using the same Lund parameters as the default 
version). For example, they overestimated the charged particle yield
and significantly underestimated the slopes of the $\pt$
spectra~\cite{Lin:2004en}. In an earlier attempt to reproduce data in
Pb+Pb collisions at LHC energies with the AMPT-SM model, the default
HIJING values for the Lund string fragmentation parameters were used
\cite{Xu:2011fi} together with the strong coupling constant
$\alpha_s=0.33$ (instead of $0.47$); there the model reasonably
reproduced the yield and elliptic flow of charged particles but
underestimated the $\pt$ spectrum (except at low $\pt$).

It was later realized that this problem of the AMPT-SM model can be
solved~\cite{Lin:2014tya} by using a very small value for the Lund
fragmentation parameter $b_L$ together with an upper limit on strange
quark productions. The AMPT-SM model can then reasonably
reproduce the pion and kaon yields, $\pt$ spectra, and elliptic flows
at low $\pt$ (below $\sim 1.5$ GeV/$c$) in central and semi-central
Au+Au collisions at the RHIC energy of  $\snn=200$ GeV and Pb+Pb
collisions at the LHC energy of 2.76 TeV~\cite{Lin:2014tya}. 
In particular, we found that $b_L=0.15$ GeV$^{-2}$ is
needed~\cite{Lin:2014tya}, which is much lower than the value used
in previous
studies~\cite{Zhang:1999bd,Lin:2000cx,Lin:2001zk,Lin:2004en,Xu:2011fi}. 
Note that, for a smaller $b_L$ value, the effective string tension 
$\kappa$, as given by~\cite{Lin:2000cx,Lin:2004en}  
\begin{equation}
\kappa \propto \frac{1}{b_L(2+a_L)},
\label{eq:kappa}
\end{equation}
is higher and thus gives a larger mean transverse momentum for the
initial quarks after string melting. 
In addition, the AMPT model assumes that the relative production of 
strange to non-strange quarks increases with the effective string 
tension \cite{Lin:2000cx,Lin:2004en}. This is because the
quark-antiquark pair production from string fragmentation in the Lund
model is based on the Schwinger mechanism \cite{Schwinger:1962tp},
where the production probability is  proportional to $\exp (-\pi
m_\perp^2/\kappa)$ at transverse mass $m_\perp$. As a result, the
strange quark suppression relative to light quarks, $\exp [-\pi
(m_s^2-m_u^2)/\kappa]$, is reduced for a higher  string tension. 
It is found that an upper limit of 0.40 on the relative production
of strange to non-strange quarks is needed for the AMPT-SM
model~\cite{Lin:2014tya}. 

Figure~\ref{fig:sm} shows the AMPT-SM results of pions and kaons for
central ($b<3$ fm) and mid-central ($b=7.3$ fm) \cite{PHENIX:2012jha}
Au+Au events at $200A$ GeV as well as central ($b<3.5$ fm) and
mid-central ($b=7.8$ fm) \cite{ALICE:2013hur} Pb+Pb events at $2.76A$ 
TeV.  Also plotted for comparisons are the corresponding data for
0-5\% and 20-30\% centralities on
dN/dy~\cite{Adler:2003cb,Bearden:2004yx,Abelev:2013vea} in panels (a)
and (d), the $\pt$ spectra at mid-rapidity for the 0-5\%  
centrality in panels (b) and (e), and $\vtwoep$ at mid-rapidity 
for the 20-30\% centrality in panels (c) and (f). 
We see good agreements between the model results and the $dN/dy$ data
in both central and mid-central events at RHIC and LHC energies. 
The value of 0.55 is used for $a_L$ at the top RHIC energy, while 
the value of 0.30 is used at the LHC energy since it gives a slightly
better fit of the ALICE data \cite{Abelev:2013vea} than the value of
0.55. We also see that the model roughly reproduces the observed $\pt$
spectra at mid-rapidity below $\sim 2$ GeV/$c$. In addition, the
AMPT-SM model roughly describes the pion and kaon elliptic flow data
on $\vtwoep$~\cite{Gu:2012br,ATLAS:2012at} at low $\pt$.  

This choice of settings for the AMPT-SM model~\cite{Lin:2014tya}  
reasonably and simultaneously reproduces the particle yield, $\pt$
spectra and elliptic flow of the bulk matter in central and
semi-central $AA$ collisions at high energies. Therefore, it enables
us to make more reliable  studies, such as the calculation of the
evolution of energy density,  effective temperatures, and transverse
flow of the parton  phase~\cite{Lin:2014tya},  and comprehensive
predictions for   Pb+Pb collisions at the top LHC energy of 5.02
TeV~\cite{Ma:2016fve}.  

\begin{figure*}[!htb]
\centering
\includegraphics [width=0.9\hsize]  {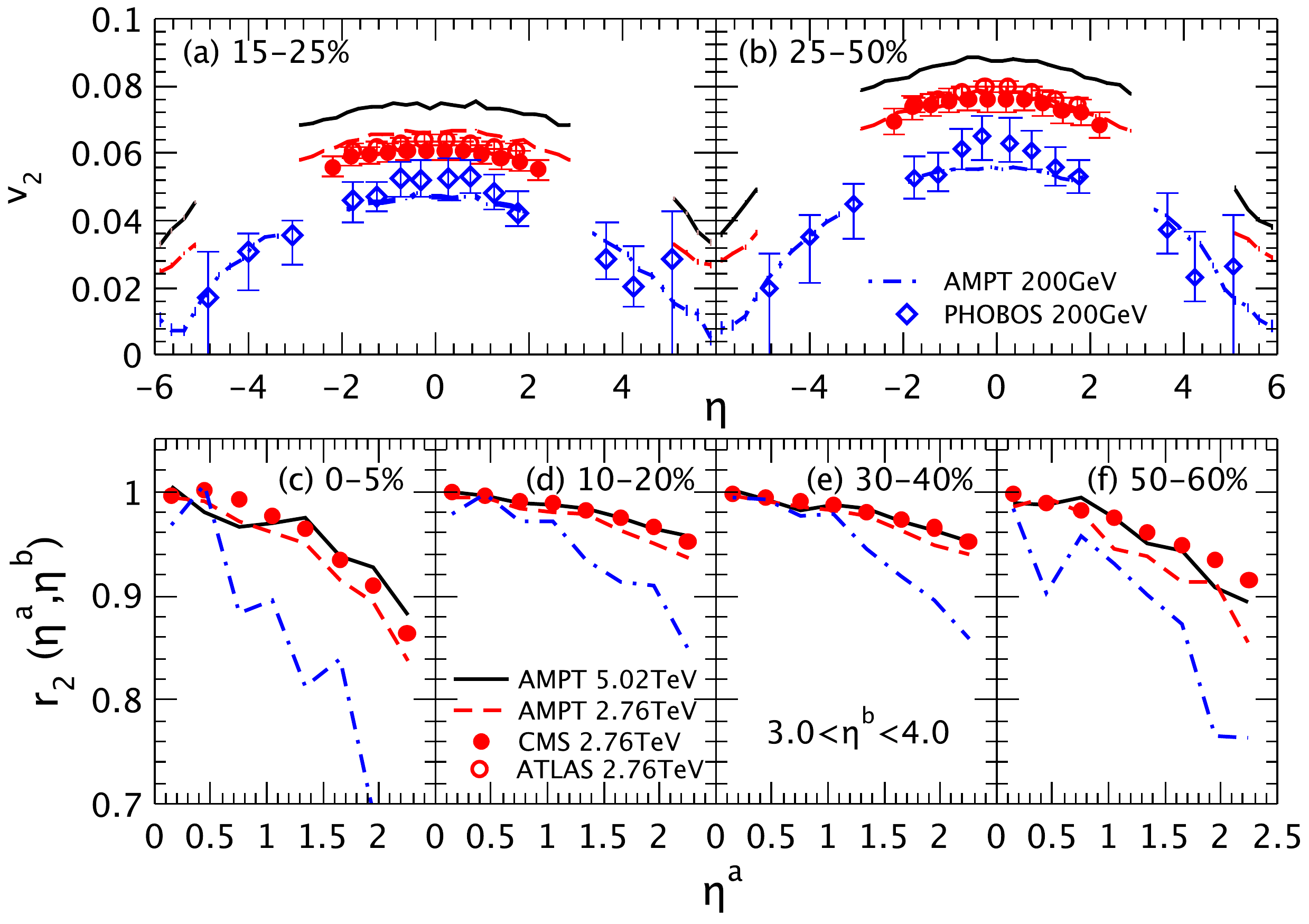}
\caption{AMPT-SM results on the $\eta$ dependence of $v_2$ in
  comparison with data for (a) the 15-25\% centrality and (b) the
  25-50\% centrality, and (c)-(f) AMPT-SM results on the factorization
  ratio $r_{2}(\eta^{a},\eta^{b})$ as functions of $\eta^a$ in
  comparison with the CMS data for different centralities.} 
\label{fig:5tev}
\end{figure*}

An example of the 5.02 TeV predictions 
from the AMPT-SM version v2.26t5~\cite{Ma:2016fve} is shown in
Fig.~\ref{fig:5tev},  where the results on the $\eta$ dependence of elliptic flow
are shown in panels (a) and (b) for two centralities and the results
on the factorization ratio $r_2(\eta^{a},\eta^{b})$ are shown in
panels (c) to (f) for four centralities. 
We see that the AMPT-SM model reasonably reproduces the observed 
$v_2(\eta)$ magnitudes and shapes at 15-25\% and 25-50\%
centralities from CMS \cite{Chatrchyan:2012ta} (filled circles) and
ATLAS \cite{Aad:2014eoa} (open circles)  for Pb+Pb collisions at 2.76
TeV and from PHOBOS \cite{Back:2004mh} (open diamonds) for Au+Au
collisions at 200 GeV. 
We also see that the AMPT results on the factorization ratio
$r_{2}(\eta^{a},\eta^{b})$ as a function of $\eta^a$ at 2.76 TeV are rather
consistent with  the corresponding CMS data
\cite{Khachatryan:2015oea},  similar to a study~
\cite{Pang:2015zrq} that used the AMPT-SM model as the 
initial condition for an ideal (3+1)D hydrodynamics. 
Furthermore, the AMPT-SM results show that the longitudinal correlation
is much suppressed in Au+Au collisions at 200 GeV 
but slightly enhanced in Pb+Pb collisions at 5.02 TeV. 
Note that the longitudinal correlation comes naturally in the AMPT-SM model
since each excited string typically produces many initial partons over
a finite $\eta$ range. 
Therefore, the initial transverse spatial geometry of the parton matter
including the event plane has a strong correlation over a finite
$\eta$ range, and through partonic and hadronic interactions, the azimuthal
anisotropies $v_n$ will then develop longitudinal correlations.

We note that the AMPT model may not be reliable at higher $\pt$, 
as indicated by Fig.~\ref{fig:sm}, since it lacks inelastic parton
collisions~\cite{Zhang:1997ej,Lin:2004en} and consequently the
radiative parton energy loss that is important for high $\pt$
partons. In addition, the string melting AMPT model up to now uses
quark coalescence to model the hadronization of all partons, while the
hadronization of high $\pt$ partons and partons far away from their
coalescing  partners should be treated differently, e.g., with
independent fragmentation~\cite{Minissale:2015zwa} or string
fragmentation~\cite{Han:2016uhh}.

\subsection{Improved quark coalescence}
\label{subsec:coal}

After parton scatterings, a spatial quark coalescence model is used to
describe the hadronization process in the AMPT-SM model. It combines a
quark with a nearby antiquark to form a meson and combines  
three nearby quarks (or antiquarks) into a baryon (or an antibaryon). 
For quarks and antiquarks in an event, the original quark coalescence
model in AMPT \cite{Lin:2001zk,Lin:2004en,Lin:2014tya,Ma:2016fve}
searches for a meson partner before searching  for baryon or
antibaryon partners. Specifically, each quark (or antiquark) has its
default coalescence partner(s), which are just the one or two
valence parton(s) from the decomposition of the quark's parent
hadron from the string melting process. 
Then for any available (i.e., not-yet-coalesced) quark (or antiquark)
that originally comes from the decomposition of a meson, the quark
coalescence model searches all available antiquarks (or quarks) and
selects the closest one in distance (in the rest frame of the
quark-antiquark system) as the new coalescence partner to form a
meson. 
After these meson coalescences are all finished, for each remaining
quark (or antiquark)  the model searches all available quarks (or
antiquarks) and selects the closest two in distance  as the new
coalescence partners to form a baryon (or an antibaryon). 
As a result, the total number of baryons in an event after quark
coalescence is the same as the total number before. 
Similarly, the quark coalescence process also conserves the number of
antibaryons and the number of mesons in an event. 

However, this separate conservation of the numbers of baryons,
antibaryons, and mesons through the quark coalescence for each
event is unnecessary, because only conserved
charges such as the number of net-baryons and the number of
net-strangeness need to be conserved. 
Therefore, we improved the coalescence
method~\cite{He:2017tla,He:2018pch} by removing the constraint that
forced the separate conservations.  
Specifically, for any available quark, the new coalescence model
searches all available antiquarks and records the closest one in
relative distance (denoted as $d_M$) as the potential coalescence
partner to form a meson. The model also searches all available quarks
and records the closest one in distance as a potential coalescence
partner to form a baryon, and then searches all other available quarks
again and records the one that gives the smallest average distance
(i.e. the average of the three relative distances among these three
quarks in the rest frame of the three-quark system, denoted as $d_B$)
as the other potential coalescence partner to form a baryon.  

In the general case where both the meson partner and baryon
partners are available, the quark will form a meson or a baryon
according to the following 
criteria~\cite{He:2017tla}: 
\begin{eqnarray}
 d_B < d_M ~r_{BM} &:& {\rm form\; a\; baryon;} \nonumber \\ 
{\rm otherwise} &:& {\rm form\; a\; meson.}
\label{eq:rbm}
\end{eqnarray}
In the above, $r_{BM}$ is the new coalescence parameter, which controls the
relative probability of a quark forming a baryon instead of forming a
meson. Note that the same coalescence procedure is also applied to all
antiquarks, and the above criteria are not needed when only the meson
partner or baryon partners can be found for a parton. 
In the limit of $r_{BM} \rightarrow 0$, there would be no
antibaryon formation at all while the minimum number of baryons would
be formed as a result of the conservation of the (positive) net-baryon
number. On the other hand, in the limit of $r_{BM} \rightarrow
\infty$, there would be almost no meson formation; more specifically,
only 0, 1, or 2 mesons would be formed depending on the remainder when
dividing the total quark number in the event by three.  
As a result, the new quark coalescence allows a (anti)quark the
freedom to form a meson or a (anti)baryon depending on the distance
from the coalescence partner(s). This is a more physical picture; for
example, if a subvolume of the dense matter is only made of quarks
which total number is a multiple of three, it would hadronize to
only baryons (with no mesons) as one would expect.

\begin{figure}[!htb]
\centering
\includegraphics[width=0.9\hsize]{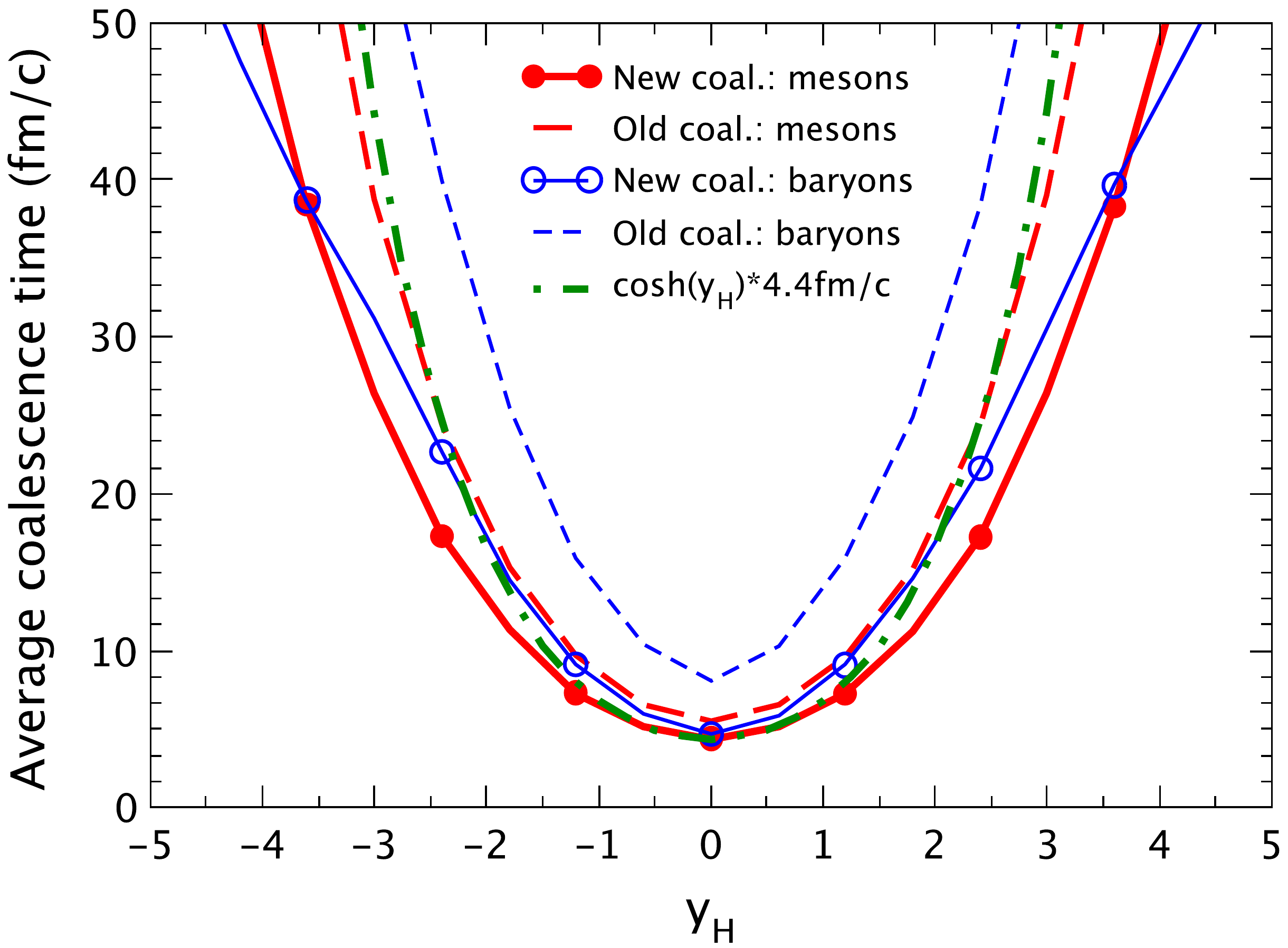}
\caption{The average coalescence time of partons in mesons and
  (anti)baryons as functions of the hadron rapidity from the new
  (curves with circles) and old (dashed curves) quark coalescence for central
  Au+Au collisions at 200 GeV; the dot-dashed curve represents a $\cosh{\rm
    y_{_H}}$ curve for comparison.} 
\label{fig:tcoal}
\end{figure}

We take central Au+Au collisions at $\snn=200 $ GeV
from the AMPT-SM model as an example to compare the old and new quark 
coalescence~\cite{He:2017tla}. 
The same  parton cross section is used so that the parton phase-space
configuration just before quark coalescence is statistically the same
for the old and new quark coalescence.
Figure~\ref{fig:tcoal} shows the average coalescence time of partons in
mesons and (anti)baryons as functions of the hadron rapidity ${\rm
  y_{_H}}$.  We see that baryons and antibaryons in the new quark
coalescence (curve with open circles) are now formed much earlier than
before. This is because the old quark coalescence tends to form
(anti)baryons late, since it  searches for meson partners before
(anti)baryon partners and a parton will be unavailable for
(anti)baryon formation when it is already used for meson formation.
In contrast, the new quark coalescence searches for the
potential meson partner and (anti)baryon 
partners concurrently and then determines the hadron type to be
formed, making the coalescence process more physical as well as more
efficient. In addition, we see that mesons in the new quark
coalescence (curve with filled circles) are also formed earlier than
before, presumably because of the improved efficiency after giving
partons the freedom to form either a meson or a (anti)baryon.
Since the plotted coalescence time is in the center-of-mass frame of
the $AA$ collision, we would expect a $\cosh{\rm y_{_H}}$ dependence if
the dense matter were boost-invariant.  The dot-dashed curve in
Fig.~\ref{fig:tcoal} represents a function that is proportional to 
$\cosh{\rm y_{_H}}$, which qualitatively agrees with our model results. 

\begin{figure}[!htb]
\centering
\includegraphics[width=0.9\hsize]{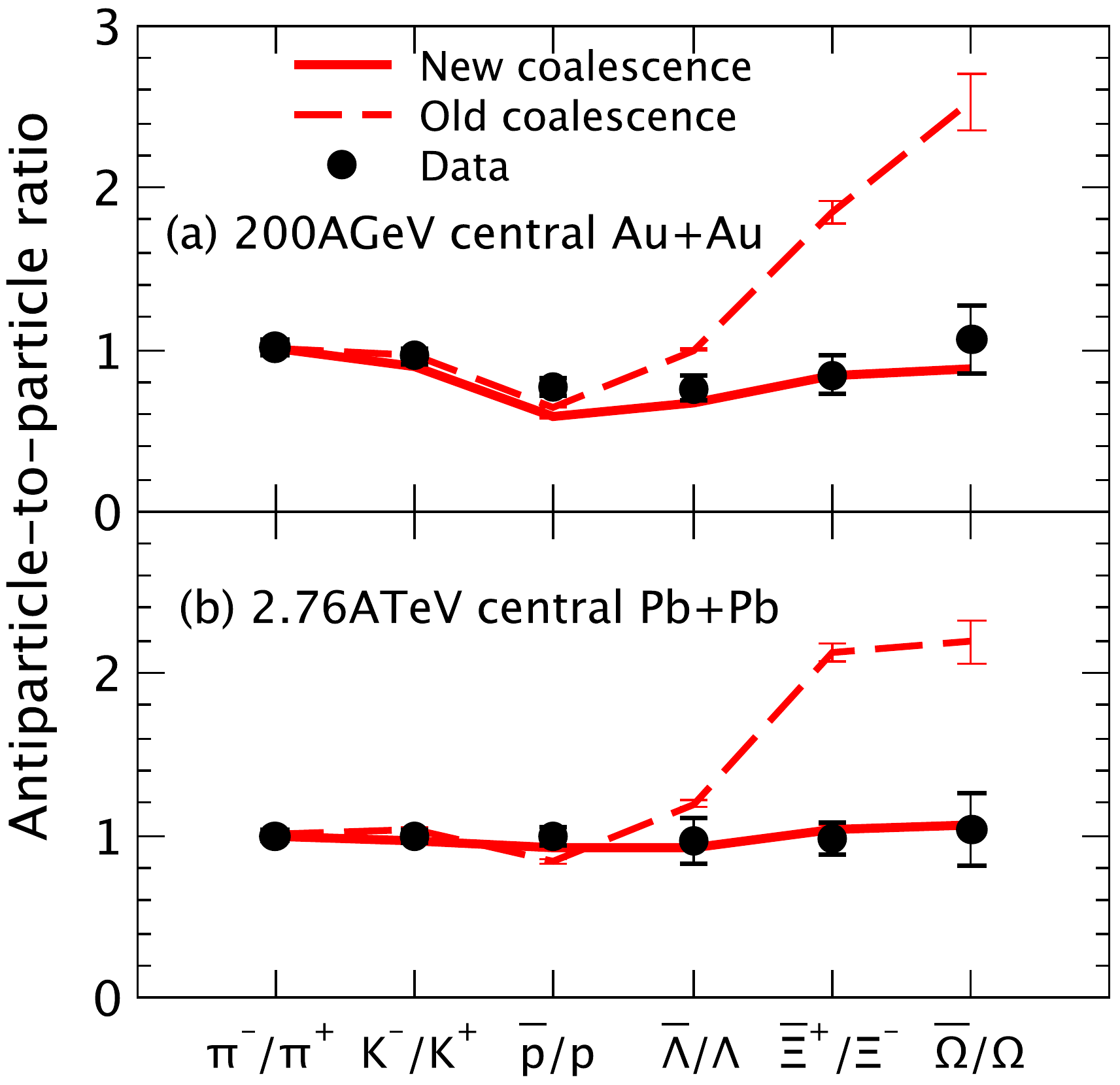}
\caption{Antiparticle-to-particle ratios around mid-rapidity for (a)
  central Au+Au collisions at 200 GeV and (b) central Pb+Pb
  collisions at 2.76 TeV from the new (solid curves) and
  old (dashed curves) quark coalescence in comparison with the
  experimental data.}
\label{fig:ratios}
\end{figure}

Therefore, the new quark coalescence is more efficient,
especially for the formation of (anti)baryons, due to the freedom of a
parton to form either a meson or a (anti)baryon. 
As a result, it leads to improvements in the descriptions of
(anti)baryon observables from the AMPT-SM
model~\cite{He:2017tla,Zhang:2018ucx,Zhang:2019bkf}. 
Figure~\ref{fig:ratios} shows the AMPT results
(with $r_{BM}=0.61$) on various
antiparticle-to-particle ratios around mid-rapidity for central Au+Au
collisions at 200 GeV~\cite{Suire:2002pa,Adams:2006ke,Abelev:2008ab} and Pb+Pb
collisions at 2.76
TeV~\cite{Abelev:2013vea,Schuchmann:2015lay,ABELEV:2013zaa} in  
comparison with the experimental data at mid-rapidity. 
Both the data and model results here are for the 0-5\%
centrality except that $\Omega$ at 200 GeV corresponds to the 0-10\%
centrality. We see that the results from the new quark coalescence
(solid curves) are generally consistent with the experimental data,
while results from the old quark coalescence (dashed curves) severely
overestimate the ratios for $\Xi$ and $\Omega$.  In addition, the
antibaryon-to-baryon ratios generally increase towards one with the
strangeness content in both the AMPT model and the data. This is
consistent with models such as the ALCOR model~\cite{Zimanyi:1999py},
which predict that these ratios are sequentially higher by a
multiplicative factor, the $K^{+}/K^{-}$ ratio.  Since the
$K^{+}/K^{-}$ ratio is usually slightly larger than one at high
energies,  we see that our results from the improved quark coalescence
agree rather well with this expectation and with the experimental
data.

On the other hand, the AMPT model with the improved quark coalescence
~\cite{He:2017tla,Zhang:2019utb} still underestimates the 
$\bar{p}/p$ ratio in central Au+Au collisions at and below 200 GeV. 
We note that quark coalescence should be augmented with other 
hadronization mechanisms such as
fragmentation~\cite{Minissale:2015zwa,Han:2016uhh} for partons that
cannot find nearby partners. This will also help avoid the
potential violation of the second law of thermodynamics 
during the hadronization process~\cite{Nonaka:2005vr}, 
where whether the entropy decreases during a phase-space quark
coalescence has been found to depend on details such as the duration
of the mixed phase, volume expansion, and resonance
decays~\cite{Greco:2007nu}. 
Also note that the $r_{BM}$ value of 0.61 is found to reasonably reproduce
the proton and antiproton yields of $AA$ collisions in the AMPT model
with the original parton distribution function and HIJING's nuclear 
shadowing~\cite{He:2017tla}, while the preferred $r_{BM}$ value is
0.53  for light ($u/d/s$) hadrons ~\cite{Zhang:2019utb, Zheng:2019alz}
and 1.0 for charm hadrons ~\cite{Zheng:2019alz} in the AMPT model with
modern parton distribution functions of nuclei.

Not only is the new quark coalescence able to describe the $dN/dy$
yields, $\pt$ spectra, and elliptic flows of pions and kaons at low
$\pt$, but it also better describes the baryon observables in general,
especially the $\pt$ spectra of (anti)baryons and antibaryon-to-baryon 
ratios for $\Xi$ and $\Omega$. It has also been shown to qualitatively
describe the near-side anticorrelation feature of baryon-baryon
azimuthal correlations observed in small systems at the 
LHC~\cite{Zhang:2018ucx,Zhang:2019bkf}.  In addition, it can be easily 
extended to include individual $r_{BM}$ factors specific to given
hadron species, e.g., to describe the enhanced multi-strange baryon
productions in nuclear  collisions~\cite{Shao:2020sqr}. The string
melting AMPT model with the  new quark coalescence thus provides a
better overall description of the bulk matter in high-energy nuclear 
collisions.

\subsection{Importance of finite nuclear thickness at lower energies}
\label{subsec:width}

For heavy ion collisions at lower energies, the thickness of the
incoming projectile and target nuclei in the center-of-mass frame
becomes larger due to the finite Lorentz contraction along the beam
directions.  Therefore, one needs to consider the finite nuclear
thickness in dynamical models of heavy ion collisions at 
lower energies, which correspond to higher net-baryon densities. 
The finite nuclear thickness increases the longitudinal width of the
created matter and thus will obviously affect the initial 
energy and net-baryon densities~\cite{Lin:2017lcj,Mendenhall:2020fil}.  
Furthermore, it will lead to a significant time
duration of the initial particle and energy production; therefore, one
cannot use a fixed proper time to describe the initial condition for
hydrodynamic-based models but use a dynamical initialization
scheme~\cite{Okai:2017ofp,Shen:2017ruz}. 

For a central collision of two identical nuclei of mass number $A$,
it takes the following time for the two nuclei to completely cross
each other in the center-of-mass frame:  
\begin{equation}
d_t=\frac{2R_A}{{\rm \sinh}\, y_{cm}} 
\label{eq:dt}
\end{equation}
in the hard sphere model of the nucleus. 
In the above, $R_A$ is the hard-sphere radius of the nucleus 
and $y_{cm}$ is the rapidity of the projectile nucleus.
For central Au+Au collisions at $\snn=50$ GeV, for
example, $d_t \approx 0.5$ fm/$c$, which is comparable to the typical
value of the parton formation time or hydrodynamics initial time when
one takes $R_A=1.12 A^{1/3}$ fm.
Therefore, one may expect the effect from finite nuclear thickness
to be significant for central Au+Au collisions at $\snn \lesssim
50$GeV, which is the focus energy range of  
the RHIC Beam Energy Scan (BES) program.

We have developed semi-analytical methods
\cite{Lin:2017lcj,Mendenhall:2020fil} to include the finite nuclear
thickness in the calculation of the initial energy density, which is
crucial in determining the initial temperature (and
net-baryon chemical potential at low energies) of the produced 
QGP. Traditionally, the Bjorken formula~\cite{Bjorken:1982qr} has been
the standard semi-analytical tool in estimating the initial energy
density in the central rapidity region right after the two nuclei pass 
each other: 
\begin{equation}
\epsilon_{Bj}(t)=\frac{1}{A_T ~t} \frac{dE_T}{dy}.
\label{eq:enebj}
\end{equation}
In the above, 
$A_T$ represents the full transverse area of the overlap volume,
and $dE_T/dy$ is the initial rapidity density of the transverse
energy at mid-rapidity, which is often
approximated with the experimental $dE_T/dy$ value in the final
state. Because the Bjorken energy density diverges as $t \rightarrow
0$, a finite value is needed for the initial time, which is often
taken as the proper formation time of the produced quanta $\tauf$. 
However, a serious limitation of the Bjorken formula results from the
fact that it neglects the finite thickness of the colliding nuclei.  
Therefore, one expects that the Bjorken formula may break down when
the crossing time is not small compared to the formation
time~\cite{Adcox:2004mh}. 

Using the semi-analytical methods that include the finite nuclear
thickness, we have calculated the initial energy density $\epsilon(t)$
averaged over the transverse area of the overlap region as a function
of time, including its maximum value $\epsilon^{\rm
  max}$~\cite{Lin:2017lcj,Mendenhall:2020fil}.  
We first considered the finite time duration of the initial energy
production but neglected the finite longitudinal
extension~\cite{Lin:2017lcj}, which enabled us to obtain explicit
analytical solutions of $\epsilon(t)$. 
Both the uniform time profile and beta time profile have been
considered, where in the uniform time profile one assumes that the
initial transverse energy at $y \approx 0$ is produced uniformly 
in time ($x$) from $\tone$ to $\ttwo$:
\ber
\frac{\dtwoetr}{dy\;dx}= {1 \over {\ttwo-\tone}} \frac{\detr}{dy},
{~\rm if~} x \in [\tone,\ttwo].
\label{uniform}
\eer
In contrast, the beta time profile assumes the following:
\ber
\frac{\dtwoetr}{dy\;dx} \propto \left [ x (d_t-x) \right ]^n
\frac{\detr}{dy}, {~\rm if~} x \in [0,d_t].
\label{beta}
\eer
Note that $n=4$ is chosen~\cite{Lin:2017lcj} from the comparison to 
the time profile of partons within mid-spacetime-rapidity in central
Au+Au collisions from the AMPT-SM model.  
In addition, for the uniform profile shown here, 
$\tone=0.29 d_t$ \& $\ttwo=0.71 d_t$ are used since they give the same
mean and standard deviation of time as the beta profile at $n=4$. 

We then considered both the finite time duration and longitudinal
extension of the initial energy production~\cite{Mendenhall:2020fil}. 
When $\tauf$ is not too much smaller than the crossing time of the two
nuclei, results from this later study~\cite{Mendenhall:2020fil} are
similar to those from the earlier study~\cite{Lin:2017lcj}. 
On the other hand, there is a qualitative difference in that the maximum
energy density $\epsilon^{\rm max}$ at $\tauf=0$ is finite
after considering both the finite duration time and longitudinal
extension~\cite{Mendenhall:2020fil}, while the
Bjorken formula  diverges as  $1/\tauf$ and the method 
that only considered the finite duration time~\cite{Lin:2017lcj}
diverges as $\ln (1/\tauf)$ at low energies but as $1/\tauf$ at 
high energies. Overall, these studies have yielded the following
qualitative conclusions: the initial energy density after considering
the finite nuclear thickness approaches the Bjorken formula at high
colliding energies and/or large formation time $\tauf$. At low
colliding energies,  however, the initial  energy density has a much
lower maximum, evolves much longer, and is much less sensitive to
$\tauf$ than the Bjorken  formula. 
Note that we have written a web interface~\cite{interface} that
performs the semi-analytical calculation~\cite{Mendenhall:2020fil} for
central $AA$ collisions, where the user can input the colliding
system, energy and the proper formation time.

\begin{figure*}[!htb]
\includegraphics [width=0.9\hsize]  {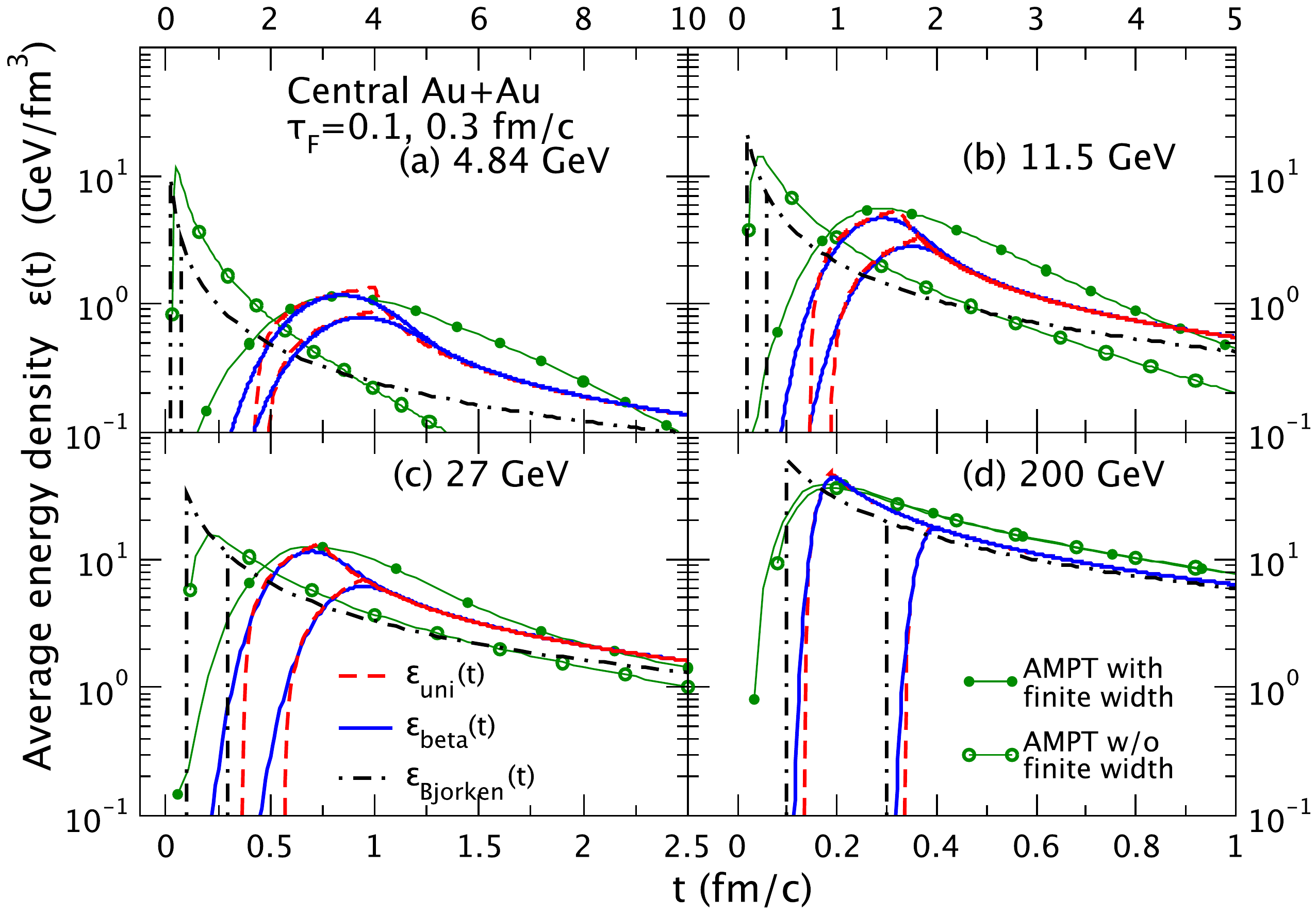}
\caption{Average parton energy densities at central spacetime-rapidity 
from AMPT with (filled circles) and without (open circles) the finite
nuclear width for central Au+Au collisions 
at (a) 4.84, (b) 11.5, (c) 27, and (d) 200 GeV; 
corresponding results for the uniform time profile (dashed
curves), the beta time profile (solid curves), and the Bjorken formula
(dot-dashed curves) at $\tauf=0.1$ \& $0.3$ fm/$c$ 
are also shown for comparison.}
\label{fig:width}
\end{figure*}

To include the finite nuclear thickness, we have modified 
the initial condition of the AMPT-SM
model~\cite{Lin:2017lcj} to specify 
in each heavy ion event the longitudinal coordinate $z_0$
and time $t_0$ of each excited string, which is then converted 
into the initial partons via string melting. 
Note that in the normal AMPT-SM
model~\cite{Lin:2001zk,Lin:2004en,Lin:2014tya,He:2017tla},  
the longitudinal coordinate $z_0$ and time $t_0$ of each
excited string in the initial state are both set to zero, which would
be correct only at very high energies. 
Figure~\ref{fig:width} shows the results on the time evolution of the
average energy density  at $\eta_s \approx 0$ for central Au+Au
collisions at four different  energies. At the high energy of 200 GeV,
the AMPT-SM results with (curves with filled circles) and without
(curves with open circles) the finite nuclear thickness are
essentially the same. This is consistent with the fact that the
Bjorken result and our semi-analytical result are also very similar
(after shifting the results in time); it also confirms the expectation
that the effect of finite nuclear thickness on the energy density can
be mostly neglected at high-enough energies. 

At lower energies, however, the AMPT results 
after including the finite nuclear thickness are very different:
the maximum energy density is lower, and the 
time evolution of the energy density (e.g., the time spent above half
the maximum energy density) is longer. These key features agree with
the semi-analytical results~\cite{Lin:2017lcj,Mendenhall:2020fil}, 
where the results from the uniform time profile and the more realistic
beta time profile are close to each other after the uniform profile is
set to the same mean and standard deviation of time as the beta
profile~\cite{Lin:2017lcj}. 
We also see from Fig.~\ref{fig:width} that the increase of the maximum 
initial energy density with the colliding energy is much faster after
including the finite nuclear thickness, which is consistent with the
analytical finding that the Bjorken formula overestimates the maximum
energy density more at lower energies \cite{Lin:2017lcj,Mendenhall:2020fil}.
In addition, we see in Fig.~\ref{fig:width} that the AMPT results are
generally wider in time; partly because the parton proper formation
time in AMPT is not set as a constant but is inversely proportional to
the parent hadron transverse mass \cite{Lin:2004en}.
Secondary parton scatterings and the transverse
expansion of the dense matter in AMPT can also cause differences from
the semi-analytical results, which do not consider such effects.
Overall, we see that the AMPT results without considering the finite
nuclear thickness are similar to the Bjorken results, while the AMPT
results including the finite thickness are similar to our semi-analytical
results. 
These results suggest that it is important to include the finite
nuclear thickness in dynamical models of  relativistic heavy ion collisions, 
especially at lower energies.

\subsection{Modern parton distribution functions in nuclei}
\label{subsec:pdf}

The initial condition of the AMPT model
is based on the HIJING two-component model~\cite{Wang:1991hta}. The primary
interactions between the two incoming nuclei are divided into two parts:  the
soft component  described by the Lund string fragmentation
model~\cite{Andersson:1983jt,Andersson:1983ia,Sjostrand:1993yb}, and the hard
component with minijet productions described by perturbative QCD through the
PYTHIA5 program~\cite{Sjostrand:1993yb}. 

The minijet differential cross sections in HIJING model can be
computed using the factorization theorem in the perturbative QCD 
framework~\cite{Eichten:1984eu} as 
\begin{equation}
\frac{d\sigma^{cd}_{\rm jet}}{d\pt^{2}dy_{1}dy_{2}} =
  K\sum_{a,b}x_{1}f_{a}(x_1, Q^{2})x_{2}f_{b}(x_2,
  Q^{2})\frac{d\sigma^{ab \to cd}}{d\hat{t}}. 
\end{equation}
In the above, $\pt$ is the transverse momentum of the produced minijet
parton, $y_{1}$ and $y_{2}$ are the rapidities of the two produced
partons $c$ and $d$, the factor $K$ accounts for higher-order
corrections to the leading order jet cross section, $x_{1}$ and 
$x_{2}$ are respectively the momentum fraction $x$ carried by the two
initial partons, $f_{a}(x_1, Q^{2})$ is the parton distribution
function (PDF) of parton type $a$ at $x=x_1$ and factorization scale
$Q^{2}$, $\sigma^{ab}$ is the parton-parton cross section for partons
$a$ and $b$, and $\hat t$ is the standard Mandelstam variable for the
minijet production subprocess.

The total inclusive jet cross section (for the production of minijet
gluons and $u/d/s$ quarks and antiquarks) is then obtained by
integrating the above differential cross section with a transverse
momentum cutoff $p_{0}$ and considering all the possible combinations
of final state parton flavors~\cite{Wang:1991hta}:
\begin{equation}
\label{eq:sigjet1}
\sigma_{\rm jet}=\frac{1}{2} \sum_{c,d} \int_{p_{0}^2}^{\hat
  s/4}d\pt^{2}dy_{1}dy_{2}\frac{d\sigma^{cd}_{\rm
  jet}}{d\pt^{2}dy_{1}dy_{2}}, 
\end{equation}
where $\hat s$ is the standard Mandelstam variable for the minijet subprocess.
We see that the minijet transverse momentum cutoff $p_{0}$ and
the parton distribution functions $f(x,Q^{2})$ are the key factors affecting
the jet cross section. The total jet cross section and the
$\sigs$ parameter that describes the soft component 
determine the nucleon-nucleon interaction cross sections in the Eikonal
formalism~\cite{Gaisser:1984pg,Pancheri:1986qg}. 
Note that $p_{0}$ is only relevant when the center-of-mass energy per
nucleon pair is higher than 10~GeV, because the jet production in the
HIJING model is switched off at $\snn<10$~GeV.  

An important ingredient needed in Monte Carlo event generators for
hadron collisions is the input parton distribution
function~\cite{Forte:2013wc,DeRoeck:2011na,Kovarik:2019xvh}. Efforts
have been made to implement various
parton distributions for phenomenological studies based on event
generators\cite{Lai:2009ne,Collins:2002ey}. The impacts of 
different parton distributions in the event generators for $pp$
collisions are found to be sizable and the key parameters in the
generators usually depend on the details of the input
PDF~\cite{Kasemets:2010sg}. Specifically, the parton distribution
function in the AMPT model affects the initial state radiation and the
minijet production within the two-component model framework. Using
modern parton distributions along with fine tuned model parameters is
required to generate reliable exclusive final states in the AMPT model.  

The HIJING~1.0 model~\cite{Wang:1991hta,Gyulassy:1994ew}
that generates the initial condition of the original AMPT model
employs the Duke-Owens parton distribution function set
1~\cite{Duke:1983gd} for the free proton. However, it is well known
that the Duke-Owens PDFs were obtained at a time when a 
large array of experimental data used in the global fittings 
for modern PDFs were not yet available~\cite{Kovarik:2019xvh}. The parton
densities at small-$x$ relevant for minijet and heavy flavor productions at
high energies are generally underestimated by the Duke-Owens
PDFs~\cite{Lin:2011zzg}. 
Therefore, we have updated the AMPT model with a modern parton
distribution function of the free nucleon (the CTEQ6.1M set
~\cite{Pumplin:2002vw}) and retuning of the relevant $p_0$ and $\sigs$
parameters~\cite{Zhang:2019utb}. Note that this update is based on the
AMPT model with the new quark coalescence~\cite{He:2017tla}.
Also note that the HIJING~2.0 model~\cite{Deng:2010mv} is a similar
update, which replaces the Duke-Owens PDFs in the HIJING~1.0 model
with the GRV PDFs~\cite{Gluck:1994uf}. 

For nuclear collisions at sufficiently high energies, results from
event generators depend on the parton distribution functions of 
the incoming nuclei. Analogous to the free nucleon case, global
analyses of the modifications of the nuclear
PDFs relative to the free nucleon PDFs have been performed by several 
groups~\cite{Eskola:2016oht,deFlorian:2011fp,Kovarik:2015cma,Hirai:2007sx,AbdulKhalek:2019mzd}. In
addition, it is natural to expect the nuclear modification to 
depend on a nucleon's position inside a nucleus. 
Therefore, the spatial dependence of nuclear parton densities are
considered~\cite{Wang:1991hta,Eskola:1991ec,Emelyanov:1999pkc,Klein:2003dj,Frankfurt:2011cs,Nagle:2010ix,Wu:2021ril}, 
and a global analysis to extract the nuclear PDFs with spatial dependence
is carried out~\cite{Helenius:2012wd} based on the
EKS98~\cite{Eskola:1998iy} and EPS09~\cite{Eskola:2009uj} fits. 
In a recent study~\cite{Zhang:2019utb}, 
we have included the spatially dependent EPS09s nuclear
modifications~\cite{Helenius:2012wd} in the AMPT model 
to replace the original HIJING~1.0 nuclear shadowing. Note that the
HIJING~1.0 shadowing is spatially dependent but independent of $Q^2$
or the parton flavor~\cite{Wang:1991hta,Lin:2004en}, similar to the
HIJING~2.0 nuclear shadowing~\cite{Deng:2010mv}.

For a proton bound in a nucleus, its parton 
distribution function of flavor $i$ can be written as
\begin{equation}
f_{i}^{p/A}(x,Q^{2}) \equiv  R_{i}^{A}(x,Q^{2})f_{i}^p(x,Q^{2}),
\end{equation}
where $f_{i}^p(x,Q^{2})$ is the corresponding PDF in the free proton. Here
$R_{i}^{A}(x,Q^{2})$ represents the spatially averaged nuclear modification
function, which typically depends on the $x$ range: the shadowing region at
small $x$, an anti-shadowing region at $x \sim 0.1$, and the EMC
effect region at $x$ close to 1. The spatial dependence of 
the nuclear modification function can be formulated as
\begin{equation}
R_{i}^{A}(x,Q^{2}) \equiv \frac{1}{A}\int d^2{\bf s}~T_{A}({\bf
  s})~r_{i}^{A}(x,Q^{2},{\bf s}),  
\end{equation}
where $T_{A}({\bf s})$ represents the nuclear thickness function at
transverse position ${\bf s}$, and $r_{i}^{A}(x,Q^{2},{\bf s})$
represents the spatially dependent nuclear modification function. 

Solid curves in Fig.~\ref{fig:PDFcompare} show the gluon density
distributions (multiplied by $x$) in the free proton from the original
and the updated AMPT model. 
The gluon density distributions  of a bound proton at the center of a 
lead nucleus from the EPS09s and HIJING nuclear modifications are also 
shown in  Fig.~\ref{fig:PDFcompare}. 
We see that in the updated AMPT model with
the CTEQ6.1M set the gluon densities are quite different from the old 
Duke-Owens set and much higher at small $x$. 
We also see that the gluon shadowing in EPS09s is much weaker than
that in the HIJING~1.0 model.
 
\begin{figure}[!htb]
\begin{center}
\includegraphics[scale=0.47]{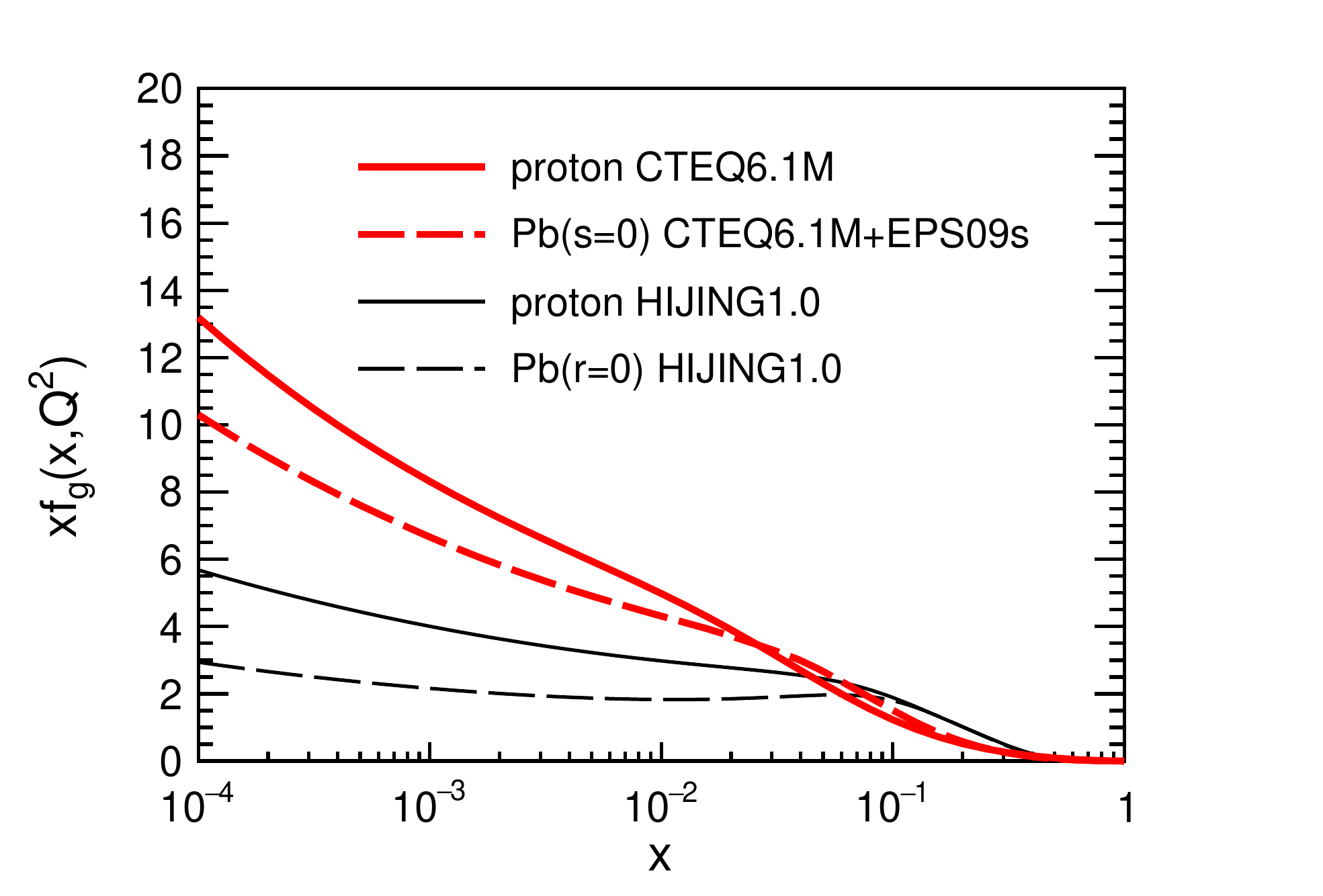}
\caption{Gluon density distributions (multiplied by $x$) in free
  proton (solid curves) and proton inside the lead nucleus (dashed
  curves) at $Q^2$=10 GeV$^2$ from the original (black) and updated
  (red) AMPT.} 
\label{fig:PDFcompare}
\end{center}
\end{figure}

As mentioned earlier, $p_{0}$ and $\sigs$ are the two key parameters in the
HIJING model that determine the total and inelastic cross sections of $pp$ and $p
\bar p$ collisions within the Eikonal formalism. In the HIJING~1.0
model that uses the Duke-Owens PDFs, constant values of $p_{0}$ = 2.0
GeV$/c$ and $\sigs$ = 57 mb are found to reasonably describe the experimental
cross sections of $pp$ and $p \bar p$ collisions over a wide energy
range~\cite{Wang:1990qp,Wang:1991hta,Gyulassy:1994ew}.  
On the other hand, when the PDFs are updated in the HIJING~2.0
model~\cite{Zhang:2019utb,Deng:2010mv}, energy-dependent
$p_{0}(s)$ and $\sigs(s)$ values are needed.

\begin{figure}[!htb]
\begin{center}
\includegraphics[scale=0.4]{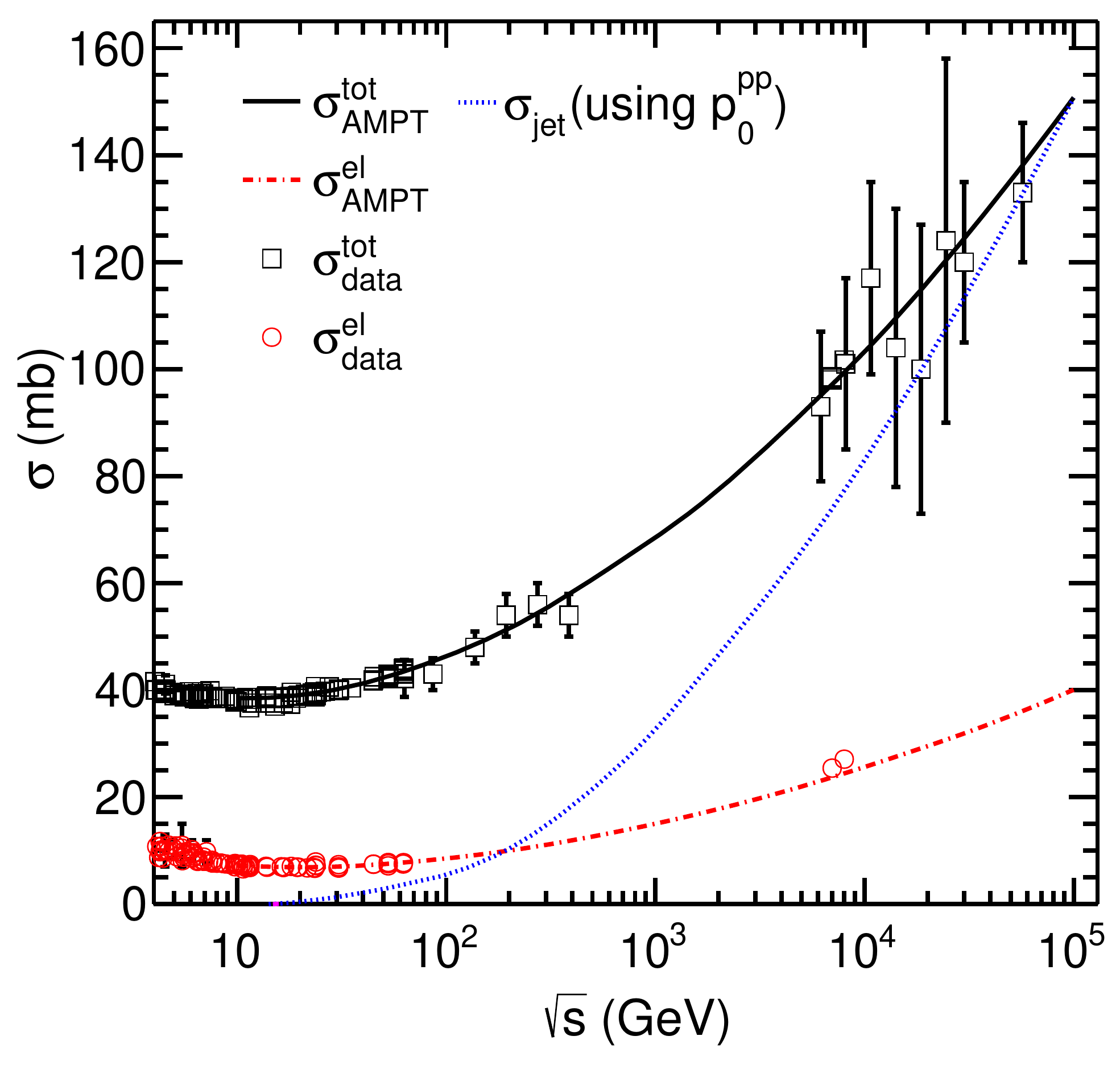}
\caption{Total and elastic cross sections versus the colliding energy
  of $pp$ collisions from the updated AMPT model (solid and dot-dashed
  curves) in comparison with the experimental data (symbols);
  $\sigma_{\rm jet}$ from the model is also shown (dotted curve).} 
\label{fig:parameterfit}
\end{center}
\end{figure}

In our work that implements the CTEQ6.1M PDF in the AMPT
model~\cite{Zhang:2019utb},  the energy dependent 
parameters  $p_{0}(s)$ and $\sigs(s)$ are determined via fitting the
experimental total and inelastic cross sections of $pp$ and $p \bar p$
collisions within the energy range $4 < \sqrt s <  10^5$ GeV, as shown
in Fig.~\ref{fig:parameterfit}. We then obtain the following 
$p_{0}(s)$ and $\sigs(s)$ functions for $pp$ collisions:
\begin{eqnarray}
\label{eq:p0sigmas}
p_{0}^{pp}(s)=&-&1.71 +  1.63 \ln(\sqrt{s})-0.256 \ln^{2}(\sqrt{s})
\nonumber \\ 
&+&0.0167 \ln^{3}(\sqrt{s}),\\
\sigs(s)=&&45.1 + 0.718 \ln(\sqrt{s})+0.144 \ln^{2}(\sqrt{s}) \nonumber \\ 
&+&0.0185 \ln^{3}(\sqrt{s}).
\end{eqnarray}
In the above, $p_{0}^{pp}$ and the center-of-mass colliding energy
$\sqrt{s}$ are in the unit of GeV$/c$ and GeV, respectively;  while
$\sigs$ is in the unit of mb. 
Note that $\sqrt{s}$ will be replaced with $\snn$ for nuclear collisions.
We also find that the updated AMPT-SM model reasonably describes the
charged particle yield and $\pt$ spectrum in $pp$ and/or $p\bar p$
collisions from $\sqrt s \sim $ 4 GeV to 13 TeV~\cite{Zhang:2019utb}.

\begin{figure*}[!htb]
\begin{center}
\includegraphics[width=0.9\textwidth]{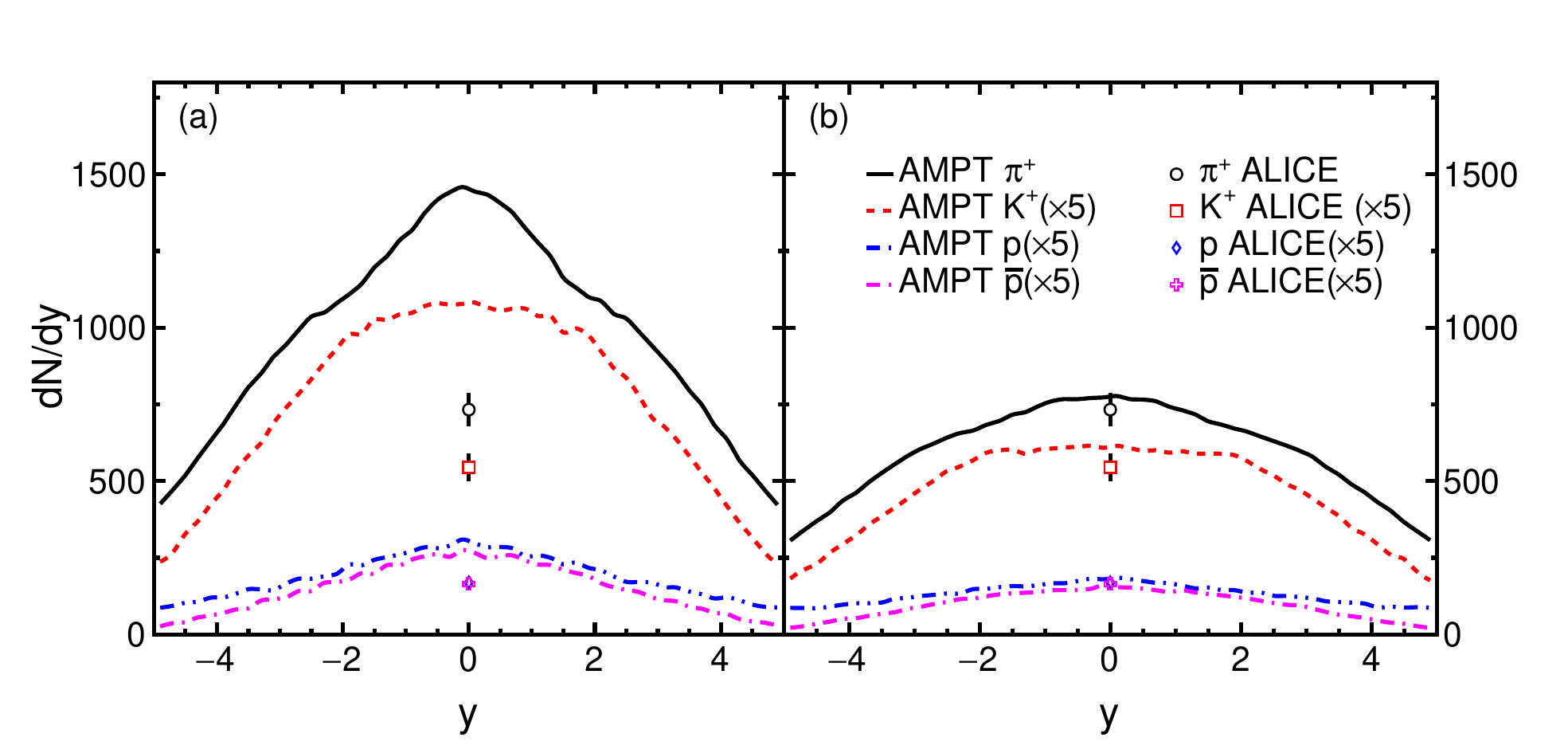}
\caption{Identified particle rapidity distributions in 0-5\% central Pb+Pb 
  collisions at $\snn=2.76$ TeV from the AMPT-SM model
  using the minijet cutoff (a) $p_0^{pp}(s)$ or (b) $p_0^{AA}(s)$ in
  comparison with the ALICE data~\cite{Adler:2003cb,Abelev:2013vea}.}
\label{fig:p0AA}
\end{center}
\end{figure*}

When we apply the above $p_{0}(s)$, $\sigs(s)$ and the EPS09s
nuclear shadowing to central $AA$ collisions at LHC energies, however, 
we find rather surprisingly that the hadron yields from the AMPT-SM
model are significantly higher than the experimental data. As shown in 
Fig.~\ref{fig:p0AA}(a), the AMPT-SM model that uses the $p_{0}^{pp}(s)$
value overestimates the final state hadron multiplicities in central
Pb+Pb collisions at $\snn=2.76$ TeV. 
Since a larger $p_0$ value would suppress the total jet cross section
and reduce the particle yields, we introduce a global scaling of the
minijet cutoff $p_0$ to make its value in central $AA$ collisions,
$p_{0}^{AA}(s)$, nuclear size dependent~\cite{Zhang:2019utb}:
\begin{eqnarray}
\label{eq:p0aa}
&&p_0^{AA}(s)=p_0^{pp}(s) A^{q(s)}, \nonumber \\ 
&&q(s)=0.0334 \ln \left ({\sqrt{s} \over E_0} \right )-0.00232 \ln^2
\left ({\sqrt{s} \over E_0} \right ) \nonumber \\ 
&&+0.0000541 \ln^3 \left ({\sqrt{s} \over E_0} \right ),
{\rm for~}\sqrt {s} \ge E_0=200{\rm~GeV.}
\end{eqnarray}
The above $q(s)$ function is determined from the fit to the overall
particle yields of central Au+Au collisions at the RHIC
energies and central Pb+Pb collisions at LHC energies (see
\cite{Zhang:2019utb} for details).
Its value is zero at $\snn \le 200$ GeV since the $p_{0}^{pp}(s)$
value works reasonably well there for central Au+Au collisions, while its
value approaches 0.16 at $\snn \sim 10^7$ GeV.  
This nuclear scaling of the minijet momentum cutoff scale $p_0$ is 
motivated by the physics of the color glass
condensate~\cite{McLerran:1993ni}, where the saturation momentum scale
$Q_s$ depends on the nuclear size as $Q_s \propto A^{1/6}$ in the
saturation regime for small-$x$ gluons in $AA$ collisions at
high-enough energies.  

As shown in Fig.~\ref{fig:p0AA}(b), the updated AMPT-SM model using
$p_0^{AA}(s)$ from the global nuclear scaling well reproduces the
identified particle yields in central Pb+Pb collisions at the LHC
energy. In addition, we find that a very small value for
the Lund $b_L$ parameter, $b_L=0.15$~GeV$^{-2}$, is needed to describe
the particle $\pt$ spectra in central $AA$
collisions~\cite{Zhang:2019utb}, similar to an earlier study~\cite{Lin:2014tya}. 
Note that recently we have generalized the minijet cutoff $p_0$ and
the Lund $b_L$ parameter with a local nuclear
scaling~\cite{Zhang:2021vvp}, as shall be discussed in
Sect.~\ref{subsec:local}, 
which would help explain why a bigger $p_0$ value but a smaller $b_L$
value  are needed for high energy $AA$ collisions than $pp$
collisions.

\subsection{Improvements of heavy flavor productions}
\label{subsec:hf}

Heavy flavors are predominantly produced from the initial hard
scatterings at early times in nuclear collisions
~\cite{Muller:1992xn,Lin:1994xma,Dong:2019byy}. 
Therefore, they are powerful observables to probe the strong
electromagnetic field created in heavy ion 
collisions~\cite{Das:2016cwd,Chatterjee:2017ahy,Nasim:2018hyw} and
transport properties of the dense
matter~\cite{He:2012df,Huggins:2012dj,Lang:2012cx,Cao:2018ews,Xu:2018gux}. 
Multiple theoretical frameworks have been developed for the description of open
heavy flavor productions in high energy $pp$ and $p$A collisions based on the pQCD
framework~\cite{Mangano:1991jk,Kniehl:2004fy,Helenius:2018uul,Cacciari:2005rk}.
Medium effects, such as those from pQCD calculations of the parton energy
loss~\cite{Djordjevic:2015hra,Xu:2015bbz} or the Langevin/Boltzmann equation
methods~\cite{Beraudo:2014boa,He:2014cla,Cao:2015hia,Uphoff:2014hza,Cao:2017hhk,Nahrgang:2013xaa,Song:2015ykw,Das:2015ana,Plumari:2017ntm} 
can be included for $AA$ collisions. 

Study of heavy flavor productions within the AMPT
model~\cite{Zhang:2005ni,Li:2018leh} has the potential to 
provide a unified model for both light and heavy flavor transport 
and improve our understanding of the non-equilibrium effects of the
QGP
evolution~\cite{He:2015hfa,Lin:2015ucn,Kurkela:2018wud,Kurkela:2018vqr}. 
In addition, using parton scatterings to model the interactions 
between heavy quarks and the evolving medium, the parton cascade
approach is able to implement any scattering angular
distribution without the need to assume small-angle scatterings.
Therefore, besides the update with modern parton distributions for
proton and nuclei as discussed in Sect.~\ref{subsec:pdf}, we have made
several significant improvements on heavy flavor productions in the
AMPT model~\cite{Zheng:2019alz}.  
First, for self consistency we include the heavy flavor cross sections
in the total minijet cross section in the HIJING two-component 
model. Second, we remove the minimum transverse 
momentum requirement ($p_0$) for initial heavy quark productions
since the heavy quark pair production cross sections from pQCD 
are already finite due to the heavy quark mass. 
These changes can be illustrated with the following modified formula
for the minijet cross section~\cite{Zheng:2019alz}:
\begin{eqnarray}
\sigma_{\mathrm  {jet}}=&& \!\!\!\! \sum_{c,d} \frac {1} {1+\delta_{cd}} 
\int_{p_0^2}^{\hat s/4} d\pt^2dy_{1}dy_{2}
\frac {d\sigma^{cd}_{\mathrm  {light}}} {d\pt^2dy_{1}dy_{2}}
\nonumber \\ 
&+&\sum_{c,d}\int_{0}^{\hat s/4}  d\pt^2dy_{1}dy_{2}
\frac {d\sigma^{cd}_{\mathrm  {heavy}}} {d\pt^2dy_{1}dy_{2}},
\label{eqn:sigjet_reform}
\end{eqnarray}
where the first term on the right hand side represents the 
cross section of light flavor ($u/d/s/g$) minijets and the second term
represents that of heavy flavors such as charm and bottom. 
Note that the factor $1/(1+\delta_{cd})$ above becomes $1/2$ for final
states with identical partons, such as $g+g \rightarrow g+g$ for
minijet gluon productions. In contrast, the original HIJING model 
uses Eq.(\ref{eq:sigjet1}) and applies the factor of $1/2$  to all
light flavor minijet production processes \cite{Wang:1991hta}, which
leads to a smaller $\sigma_{\rm jet}$ than
Eq.(\ref{eqn:sigjet_reform}) (at the same $p_0$). 
As a result, an increase of the $p_0$ value as shown below is
needed~\cite{Zheng:2019alz} for Eq.(\ref{eqn:sigjet_reform}) to
describe the experimental data on total and inelastic cross sections
of $pp$ and $p \bar p$ collisions shown in
Fig.~\ref{fig:parameterfit}:  
\begin{eqnarray}
\label{eq:p0pp_charm}
p_0^{pp}(s)&=&-1.92 +  1.77 \ln(\sqrt{s})-0.274 \ln^{2}(\sqrt{s}) \nonumber \\ 
&&+0.0176 \ln^{3}(\sqrt{s})
\end{eqnarray}
with $p_0$ in GeV$/c$ and $\sqrt{s}$ in GeV. 

\begin{figure}
\includegraphics[width=0.54\textwidth]{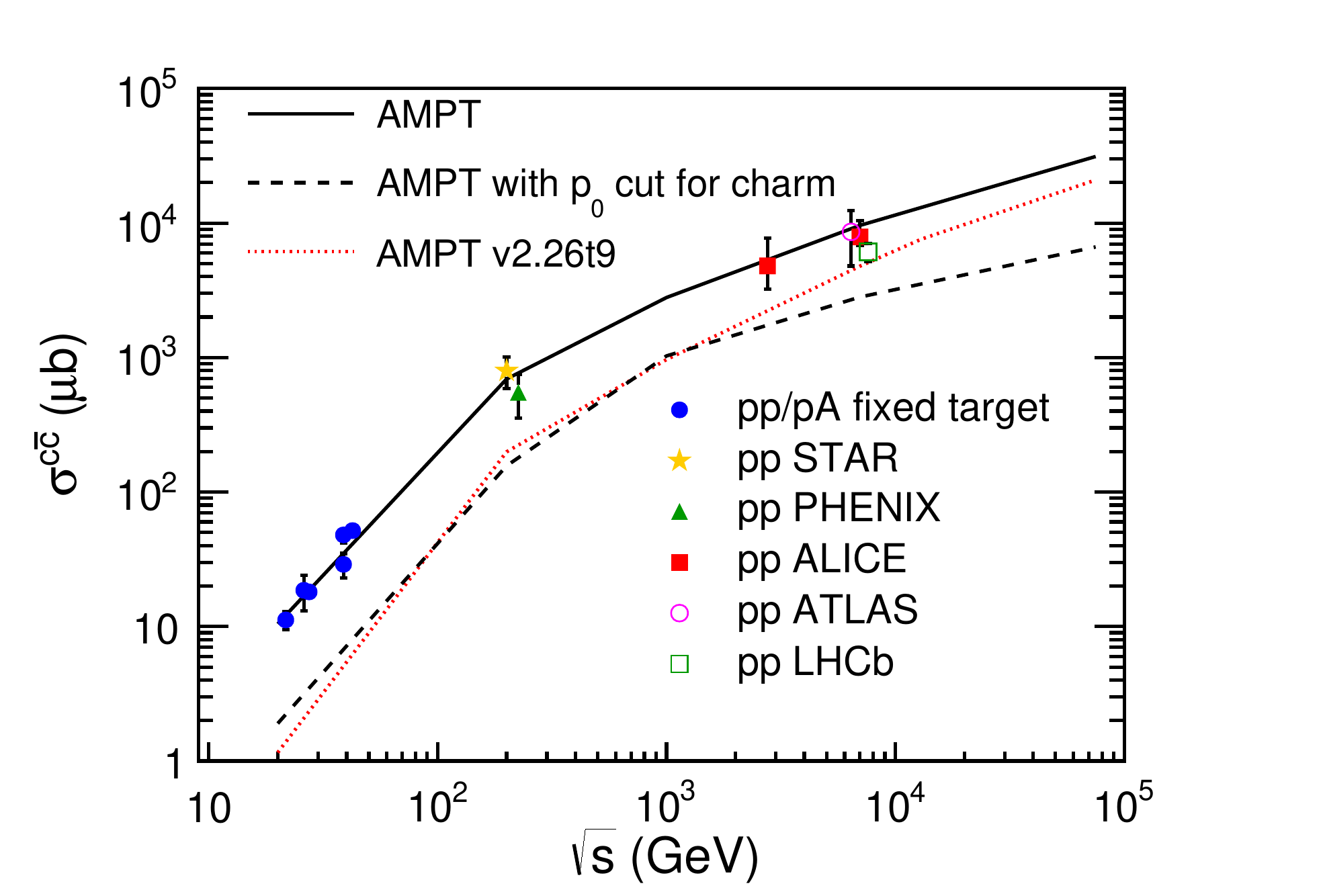}
\caption{Total cross sections of charm-anticharm quark pairs from
the AMPT model for $pp$ collisions in comparison with the world
data~\cite{Lourenco:2006vw,Adare:2010de,Adamczyk:2012af,Aaij:2013mga,Aad:2015zix,Abelev:2012vra,Acharya:2017jgo}
as functions of the colliding energy.}
\label{fig:sig_ccbar_energy_pp}
\end{figure}

The total $c\bar{c}$ cross section for $pp$ collisions from the
updated AMPT model (solid curve) is shown in
Fig.~\ref{fig:sig_ccbar_energy_pp}  versus the colliding energy in
comparison with the available world data. 
We see that the updated AMPT model can well describe the data in $pp$
collisions at various collision energies. The original AMPT model
(dotted curve), however, significantly underestimates the charm quark  
yield, especially at low energies.
The enhanced charm quark production in the updated model is largely
due to the removal of the $p_0$ cut, since the charm quark cross
section is much lower when charm quarks 
in the updated AMPT model are required to have 
a transverse momentum above $p_0$ (dashed curve).

Figure~\ref{fig:pp_cquark} shows the charm quark rapidity and
transverse momentum distributions from the AMPT model for $pp$
collisions at $\sqrt{s}=200$ GeV and 7 TeV. 
Note that the charm quark or hadron results shown in this section have
been averaged over those for particles and the corresponding 
antiparticles. As shown in Fig.~\ref{fig:pp_cquark}(a), 
the charm quark yield in the updated AMPT model is
found to be significantly higher than that in the original AMPT model
over the full rapidity range at both RHIC and LHC energies. 
From the charm quark $\pt$ spectra at
mid-rapidity shown in Fig.~\ref{fig:pp_cquark}(b), we see that the
removal of the $p_0$ cut for charm quarks mostly enhances charm quark
productions at low $\pt$.  We also see that results from the 
original AMPT model (dotted curve) and the updated AMPT model that
includes the $p_0$ cut (dashed curve)  are similar, partly because a
$p_0$ cut ($2$ GeV$/c$) on the charm quark production is also
used in the original AMPT model.

\begin{figure*}[hbt]
\centering
\includegraphics[width=0.49\textwidth]{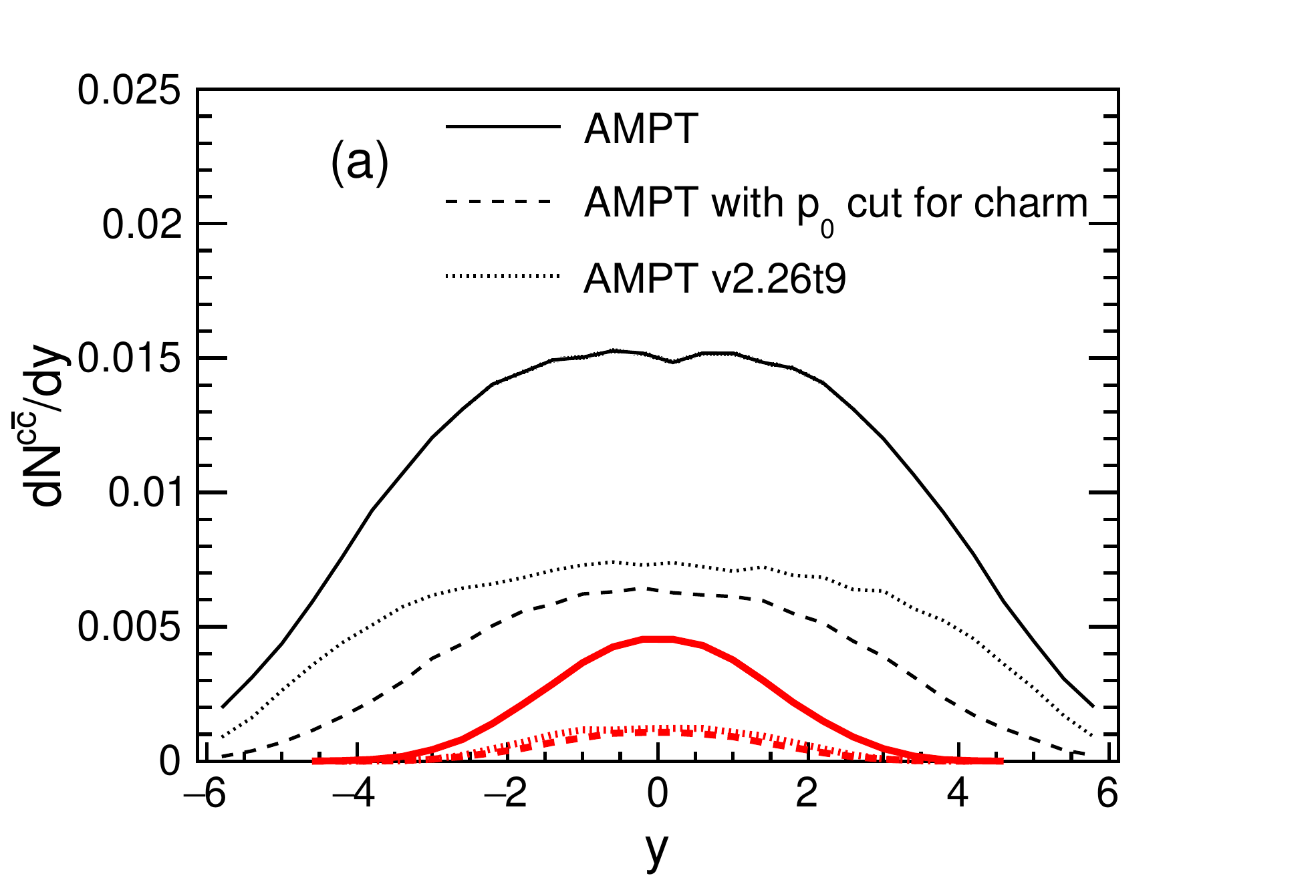}
\includegraphics[width=0.49\textwidth]{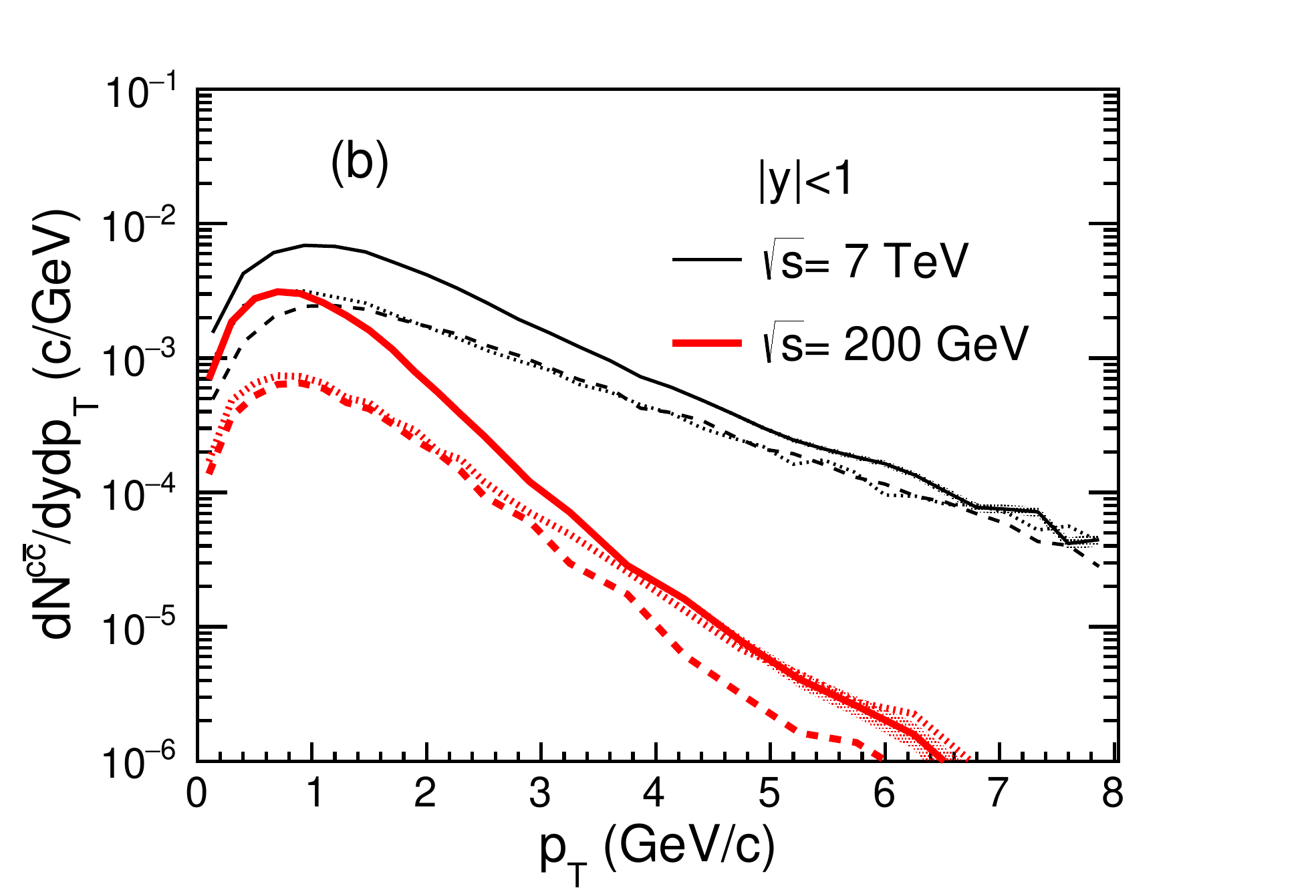}
\caption{Charm quark (a) rapidity distributions
and (b) $\pt$ spectra around mid-rapidity in $pp$ collisions at $\sqrt{s}=200$
GeV (red curves) and 7 TeV (black curves) from the AMPT model. 
The shaded band represents statistical errors.}
\label{fig:pp_cquark}
\end{figure*}

In $AA$ collisions, heavy quark production is subject to additional
medium-induced initial state and final state effects. Within the AMPT
model, initial state effects include the nuclear modification of the
parton distribution functions in nuclei, while the final state effects
are mostly treated with parton elastic rescatterings in the parton
cascade~\cite{Zhang:1997ej}. Figure~\ref{fig:dndy_ccbar_energy_AA}
shows the charm quark yield at mid-rapidity for 0-10\%  central Au+Au
or 0-10\%  central Pb+Pb collisions at different energies. 
We see that the EPS09s nuclear modification leads to 
an enhancement of the charm quark yield in central $AA$ collisions 
at lower energies but a suppression at high energies.
This is expected due to the anti-shadowing feature at
large $x$ and the shadowing feature at small $x$ in the nuclear
modification functions. We also see that the result from the updated
AMPT model (solid curve) is in good agreement with the charm quark 
data, which is obtained for 0-10\% central Au+Au collisions at $\snn=200$ GeV
by scaling the STAR $pp$ charm quark cross section data
with the number of binary collisions~\cite{Adam:2018inb,Xie:2018thr}. 
Similar to the results for $pp$ collisions, the updated AMPT model
gives a significantly higher charm quark yield at mid-rapidity in
central $AA$ collisions compared to the original AMPT model. 

\begin{figure}
\includegraphics[width=0.54\textwidth]{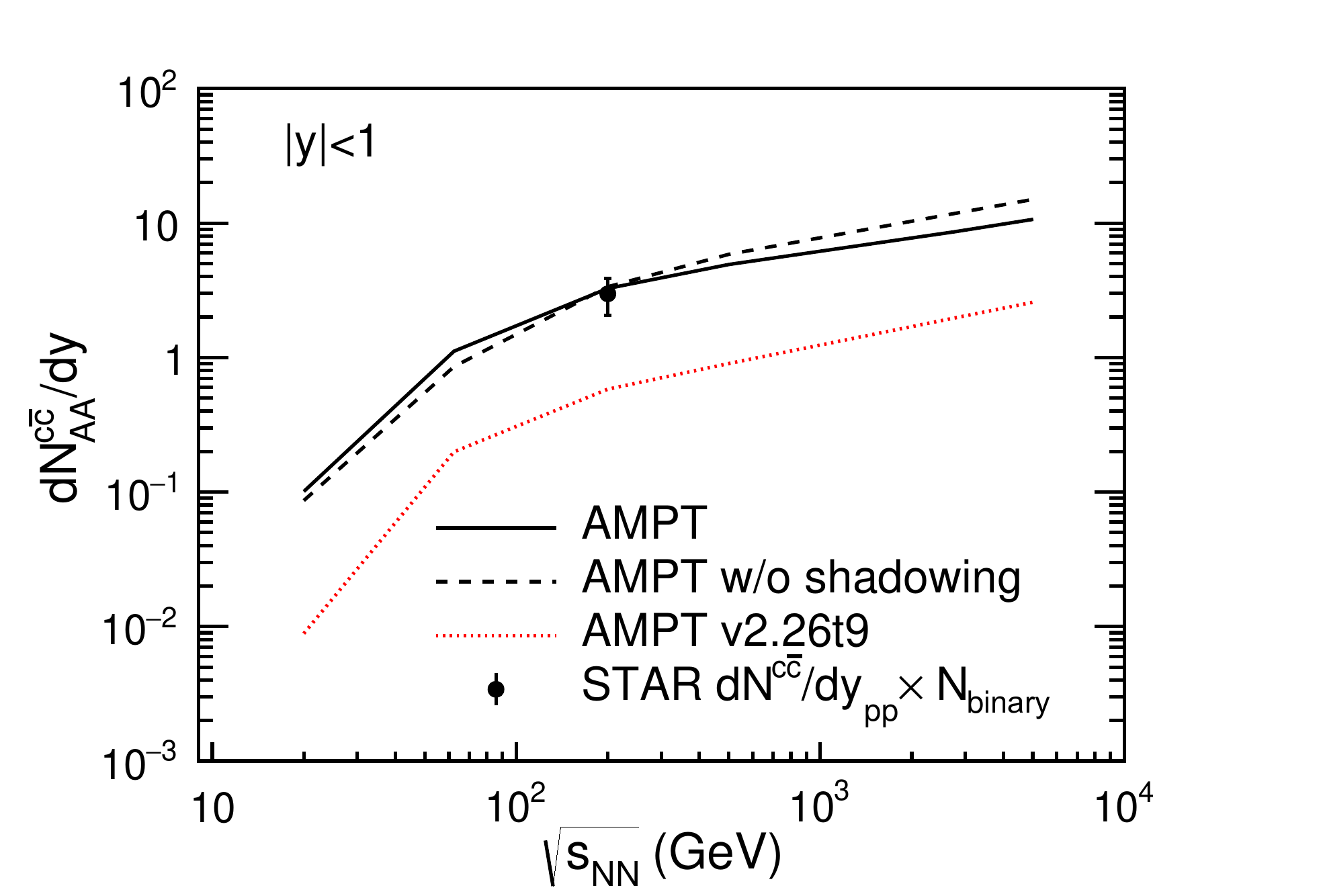}
\caption{$dN/dy$ of charm quark pairs around mid-rapidity
from the AMPT model for 0-10\% central Au+Au collisions at RHIC energies and
Pb+Pb collisions above RHIC energies as functions of the colliding
energy in comparison with the STAR data.}
\label{fig:dndy_ccbar_energy_AA}
\end{figure}

The open heavy flavor hadron species formed by quark coalescence
include charm and bottom hadrons with all possible charges. 
To reproduce the observed vector to pseudo-scalar meson ratios 
of open heavy flavors in $pp$ collisions, 
we fit the relative probability of forming primordial vector 
versus pseudo-scalar heavy mesons in the quark coalescence 
model, e.g., the ratio is set to 1.0 for the primordial $D^*/D$ and
$B^*/B$ ratios~\cite{Zheng:2019alz}, 
instead of using only the invariant mass of the coalescing partons in
the original AMPT-SM model~\cite{Lin:2004en}. 
Note that only the primordial ratios right after coalescence are 
specified with these parameters, not the vector to pseudo-scalar
meson ratios in the final state which include effects from resonance decays.
In addition, in the new quark coalescence model~\cite{He:2017tla} that
is used in this heavy flavor work~\cite{Zheng:2019alz}, the relative
probability for a quark to form a baryon instead of a meson is 
determined by the $r_{BM}$ parameter as shown in Eq.(\ref{eq:rbm}). 
In our earlier work that updated the AMPT model with modern
PDFs~\cite{Zhang:2019utb}, the $r_{BM}$ value for light flavor
$(u/d/s)$ hadrons is set to 0.53, which value is also used here. 
On the other hand, we set the $r_{BM}$ value for heavy flavor hadrons
to 1.0, because using the light flavor value would lead to
too few charm baryons (by a factor of $\sim 4$) compared to the
experimental data in $pp$ or $AA$ collisions. 
In principle, the $r_{BM}$ value for charm hadrons depends on
properties such as the number and masses of available charm baryon 
states versus charm meson states. The necessity of using a higher
$r_{BM}$ value for charm is consistent with the assumption that there
are more charm baryon states than charm meson states compared to the
light flavor sectors~\cite{He:2019tik}.

\begin{figure*}[hbt]
	\centering
	\includegraphics[width=0.49\textwidth]{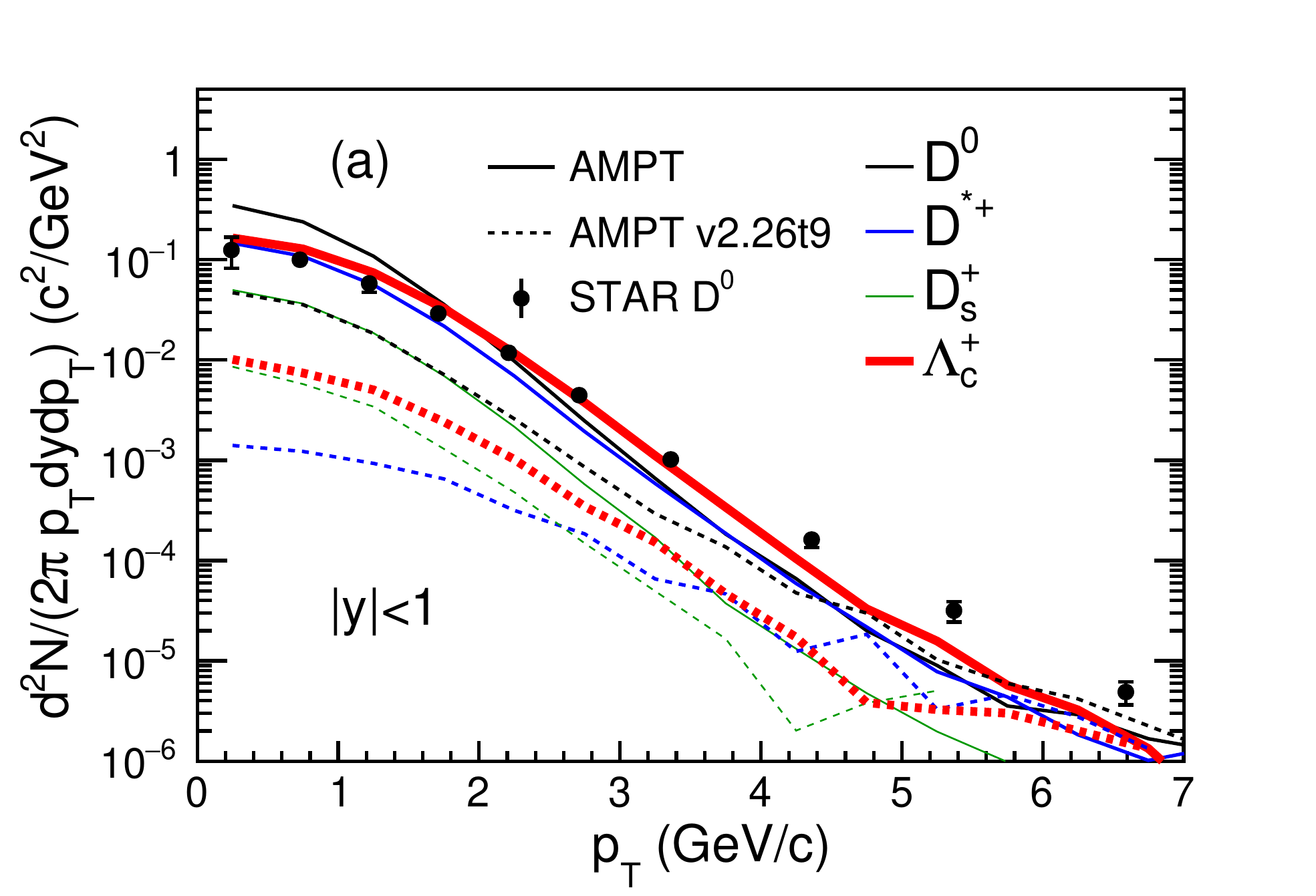}
	\includegraphics[width=0.49\textwidth]{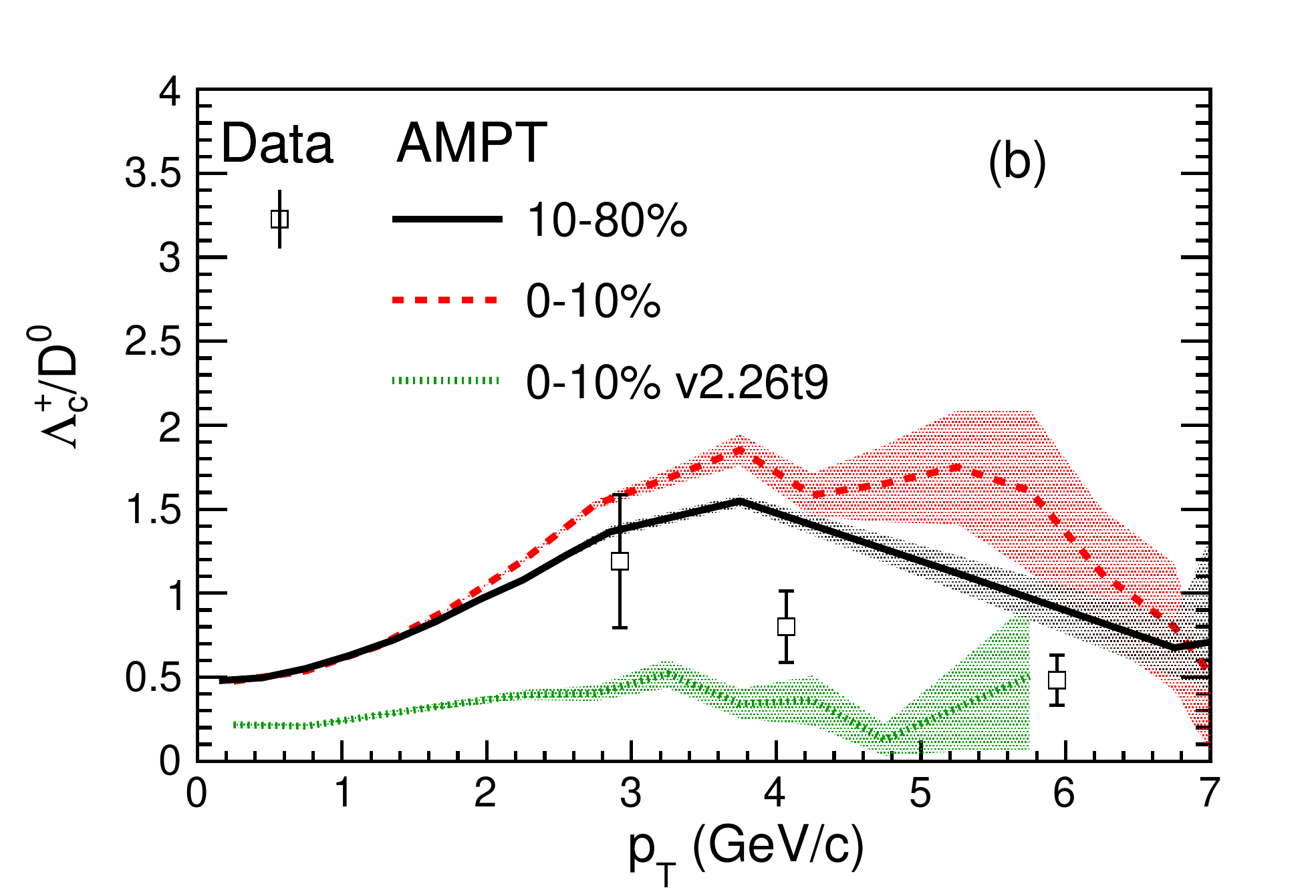}
	\caption{(a) $\pt$ spectra of open charm hadrons in central
          Au+Au collisions at $\snn=200$ GeV and (b) the 
          $\Lambda_c/D$ ratios versus $\pt$ for two centralities of
          Au+Au collisions at 200 GeV from the AMPT model in comparison
          with the STAR data~\cite{Adam:2018inb,Adam:2019hpq}.}
	\label{fig:AA_charm_hadron_RHIC}
\end{figure*}

After the improvements on heavy flavor productions, we find that the
updated AMPT model~\cite{Zhang:2019utb} can well describe the  
yields and $p_{\rm T}$ spectra of open charm hadrons
including $D$, $D^*$, $D_s$ and $\Lambda_c$
in $pp$ collisions at different energies. 
The updated model also describes the charm data
in central $AA$ collisions much better than the original AMPT model. 
However, the updated AMPT model still does not well describe the charm
hadron productions in $AA$ collisions~\cite{Zheng:2019alz}. 
As shown in Fig.~\ref{fig:AA_charm_hadron_RHIC}(a), the updated AMPT
model overestimates the $D^0$ yield at low $\pt$ but underestimates it 
at $\pt$ above 2.5 GeV/$c$ when compared to the STAR
data for 0-10\% Au+Au collisions at $\snn=200$
GeV~\cite{Adam:2018inb}. 
Compared to the original AMPT model, the charm hadron yield 
from the updated AMPT model is significantly enhanced at low $\pt$,
similar to the results at the parton level shown in
Fig.~\ref{fig:pp_cquark}(b). We also find that the charm baryon to
meson ratio ($\Lambda_c/D$)  in 0-10\% Au+Au collisions at 200 GeV is
much larger in the updated AMPT model (dashed curve) than the original
AMPT model (dotted curve), as shown in
Fig.~\ref{fig:AA_charm_hadron_RHIC}(b).  
Compared to the STAR data for 10-80\% Au+Au collisions at 200
GeV~\cite{Adam:2019hpq}, result of the $\Lambda_c/D$ ratio
versus $\pt$ from the updated AMPT model is somewhat higher. 
We also see that this ratio from the AMPT model is slightly lower in
the 10-80\% centrality than the 0-10\% centrality for Au+Au collisions
at 200 GeV.

We note that only elastic parton scatterings are included in the
AMPT-SM model; therefore, the model is only applicable in the
region where the effect of parton radiative energy loss is small. 
Studies have suggested that the elastic collisional energy
loss could be dominant for charm hadrons below $\pt \sim 5-6$
GeV$/c$ in Au+Au collisions at $\snn$ = 200 GeV or below $\pt \sim 15$
GeV$/c$ in Pb+Pb collisions at $\snn$ =
2.76TeV~\cite{Song:2015ykw}. Therefore, the charm results from the
AMPT model are not reliable at $\pt$ higher than these values. 
Also note that the charm $\pt$ spectra are affected by the charm quark
scattering cross section and its angular distribution in ZPC. The AMPT
model currently uses the $g+g  \rightarrow g+g$ cross section for 
scatterings of all parton flavors, where flavor-dependent cross
sections and angular distributions should be used for the parton
scatterings. In addition, there is still a large uncertainty on the nuclear
shadowing of gluons~\cite{Helenius:2012wd}, which has not been fully
explored in the AMPT model. 
Furthermore, hadronic scatterings of heavy flavor
hadrons~\cite{Lin:1999ad,Lin:1999ve,Lin:2000jp,Sibirtsev:2000aw,Bierlich:2021poz} 
have not been included in the AMPT model except for the decays of 
heavy hadron resonances.
Future developments of the AMPT model are expected to improve its
description of heavy flavor productions in $AA$ collisions. 

\subsection{System size dependence under local nuclear scaling}
\label{subsec:local}

The system size dependence of observables can be useful to uncover the
transition of certain phenomena in nuclear collisions, such as 
the onset of collectivity and whether it comes from initial state
momentum correlations~\cite{Dusling:2017dqg,Mace:2018vwq} or final
state
interactions~\cite{He:2015hfa,Lin:2015ucn,Weller:2017tsr,Kurkela:2018ygx,Kurkela:2019kip}.  
It is known from multiple studies that certain key parameters in the
initial condition of the AMPT model for $AA$ collisions need to be
different from their values for $pp$ collisions to reasonably describe the
data~\cite{Lin:2000cx,Lin:2004en,Xu:2011fi,Lin:2014tya,He:2017tla,Zhang:2019utb}. 
First, the Lund $b_L$ parameter in the symmetric string fragmentation
function~\cite{Andersson:1983jt,Andersson:1983ia}, as shown in
Eq.(\ref{eq:lund}), for large collision
systems needs to be significantly smaller than its value for $pp$
collisions. An earlier study has also shown that a constant $b_L$ can
not describe the centrality dependence of $\langle \pt \rangle$ in
heavy ion collisions~\cite{Ma:2016fve}, where the system size
dependence of the Lund fragmentation parameters was suggested as a
possible solution. Note that similar frameworks for the system size
dependence have been implemented in the string fragmentation 
model~\cite{Andersson:1991er,Bierlich:2015rha,Bierlich:2014xba,Fischer:2016zzs,Bierlich:2017sxk}. Second, we have found in earlier developments of the AMPT
model~\cite{Zhang:2019utb,Zheng:2019alz} that the minijet transverse
momentum cutoff $p_0$ for central Pb+Pb  collisions at the LHC
energies needs to be significantly higher than that for $pp$
collisions at the same energy. These observations suggest that the
above two parameters should be related to the size of the colliding
system to provide better initial conditions for the AMPT model. 

Therefore, we have recently proposed~\cite{Zhang:2021vvp} that the
$b_L$ and $p_0$ parameters in AMPT can be considered as local
variables that  depend on the nuclear thickness functions of the two
incoming nuclei. This prescription allows us to use the parameter
values obtained for $pp$ collisions and the local nuclear scaling
relations to  obtain the values for $AA$ collisions; the model would
then describe the system size and centrality dependences of  nuclear
collisions self-consistently.  

In the Lund string model~\cite{Andersson:1983jt,Andersson:1983ia}, 
the symmetric fragmentation function is given by 
Eq.(\ref{eq:lund}). 
The average squared transverse momentum of massless
hadrons is related to the Lund fragmentation parameters $a_L$
and $b_L$ as~\cite{Lin:2004en}
\begin{equation}
\langle \pt^{2}\rangle=\frac{1}{b_L(2+a_L)}.
\end{equation}
Consequently, the $\langle \pt\rangle$ of both partons after string
melting and the final hadrons are significantly affected by the value
of $b_L$. Since the mean transverse momentum of initial partons in
heavy ion collisions is expected to be higher in larger systems due to
the higher initial temperature, we expect the $b_L$ value to decrease
with the system size. Note that the string tension is believed to be
larger in a denser
matter~\cite{Biro:1984cf,Tai:1998hd,Bierlich:2014xba,Fischer:2016zzs}, 
thus a decrease of $b_L$ with the system size is consistent with the
expectation of a stronger color field and thus a higher string
tension $\kappa$ since $\kappa \propto 1/b_L$\cite{Lin:2004en}  
as shown in Eq.(\ref{eq:kappa}).

We propose that $b_L$ depends on the local transverse position of the
corresponding excited string inside the nucleus in each
event~\cite{Zhang:2021vvp}. Specifically, we assume that $b_L$ scales
with the local nuclear thickness functions in a general $AB$ collision as 
\begin{eqnarray}
\label{eq:bl}
b_L(s_A,s_B,s)=\frac {b_L^{pp}} {\left [ \sqrt {T_A(s_A) T_B(s_B)}/T_p \right ]^{\beta(s)}}.
\end{eqnarray}
In the above, $b_L^{pp}$ is the value for $pp$ collisions (chosen to be 0.7
GeV$^{-2}$ based on the fit of $\langle \pt \rangle$ data), $s$
represents the square of the center-of-mass 
collision energy per nucleon pair, $T_A(s_A)=\int\rho_{A}(s_A,z)dz$ is the
nuclear thickness function at the transverse distance $s_A$ from the center of
nucleus $A$ determined with the Woods-Saxon nuclear density
profiles~\cite{Eskola:1998iy}, and $T_p$ is the average value of the
effective thickness function of the proton (taken as 0.22
fm$^{-2}$). Note that $T_p$ is used instead of $T_A(s_A)$ or
$T_B(s_B)$ when the projectile or the target is proton or when
$T_A(s_A)$ or $T_B(s_B)$ from the nucleus is smaller than the $T_p$ 
value. Although different mathematical forms from that of
Eq.(\ref{eq:bl}) can be used in the local scaling relation, 
our study~\cite{Zhang:2021vvp} shows that the geometric scaling
form (i.e., using the geometric mean of the two nuclear
thickness functions) generally works better than the arithmetic form. 
We note that a systematic Bayesian analysis based on the TRENTo
initial condition~\cite{Moreland:2014oya} with a hybrid model found
that the geometric form for the local nuclear scaling is preferred by
the experimental data~\cite{Bernhard:2016tnd}. 

The exponent function $\beta(s)$ describes the energy dependence of
the local nuclear scaling of $b_L$. From the fits to charged
particle $\langle \pt \rangle$ data in the most central Au+Au collisions at
RHIC energies and most central Pb+Pb collisions at LHC energies, 
it is parameterized as
\begin{eqnarray}\label{eq:betas}
\beta(s)=0.620+ 0.112 \ln \left (\frac{\sqrt{s}}{E_0} \right ) \Theta (\sqrt{s}-E_0),
\end{eqnarray}
where $E_0=200$ GeV and $\Theta(x)$ is the unit step function. 
The fitted $\beta(s)$ function is shown in Fig.~\ref{fig:blp0ab}(a)
(dashed curve), which is a constant at RHIC energies but grows rapidly
at LHC energies.  
Note that $\beta=1$ at high energies (dotted line) may be a
``natural'' limit for Eq.\eqref{eq:bl} if we imagine that all local
strings fully overlap so that the string tension adds up. That
would give $b_L  \propto 1/T_A(s_A)$ for central $AA$ collisions, 
where $T_A(s_A)$ is proportional to the local number of participant
nucleons or excited strings integrated over the longitudinal length. 

\begin{figure}[!htb]
\includegraphics[scale=0.43]{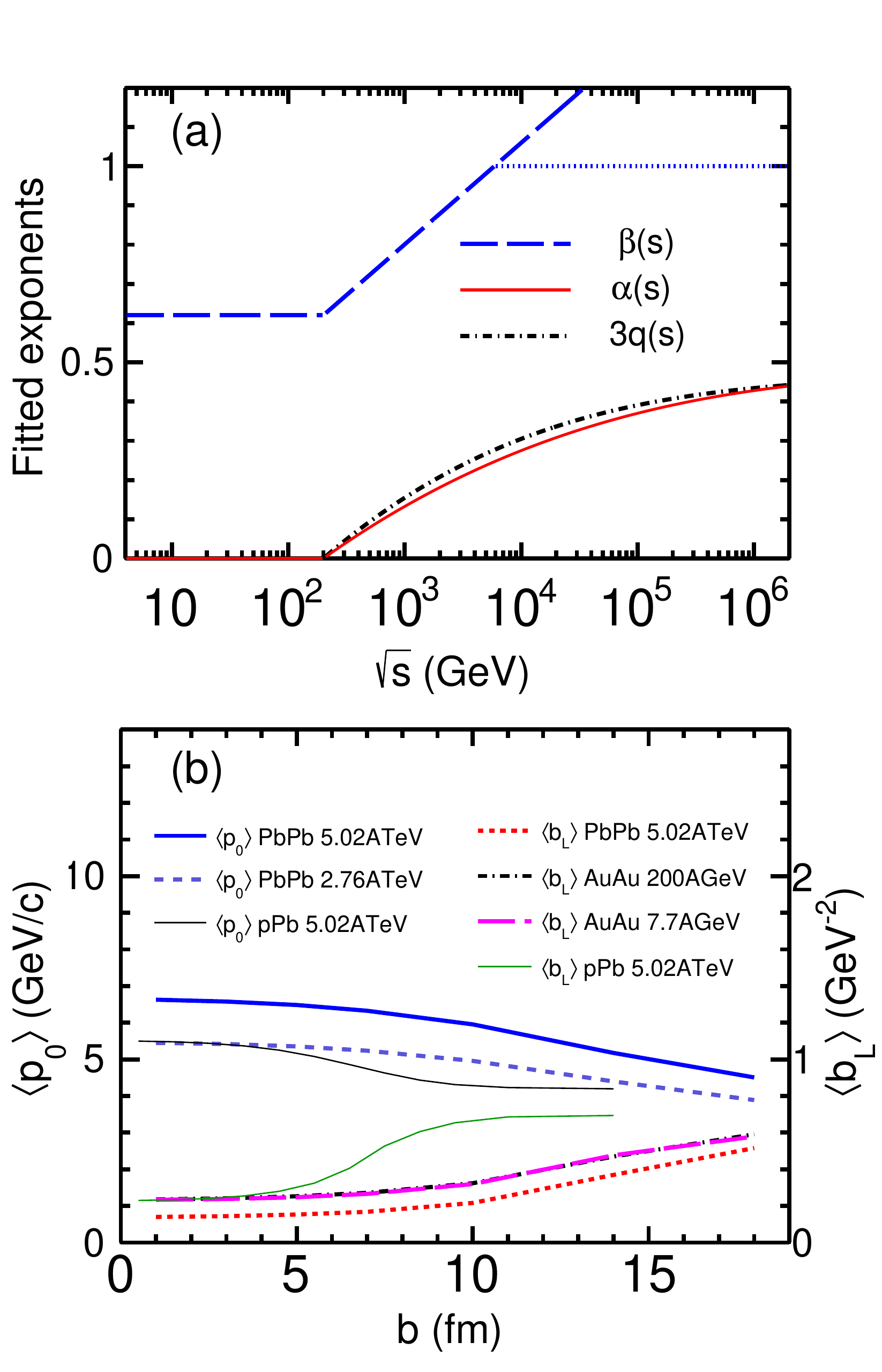}
\caption{(a) Fitted exponent functions $\alpha(s)$, $\beta(s)$, and
  $3q(s)$ versus the  center-of-mass energy per nucleon pair
  $\sqrt{s}$.  (b) Average $p_0$ and $b_L$ values versus the impact
  parameter of $p$Pb, Au+Au and Pb+Pb collisions at several energies.} 
\label{fig:blp0ab}
\end{figure}

Figure~\ref{fig:blp0ab}(b) shows the $b_L$ value averaged over the
overlap volume versus the impact parameter for Pb+Pb and $p$Pb 
collisions at $5.02A$ TeV and Au+Au collisions at two RHIC energies. 
We see that $\langle b_L\rangle$ for Pb+Pb collisions at the LHC
energy is lower than that for Au+Au collisions at RHIC energies, which
corresponds to a larger string tension due to the larger value of the
exponent $\beta(s)$ at LHC energies. 
On the other hand, the impact parameter dependences of  $\langle b_L
\rangle$ at different RHIC energies are essentially the same since 
$\beta(s)$ is a constant within that energy range. 
For $p$Pb collisions at $5.02A$ TeV, its $\langle b_L \rangle$ is
higher than that in Pb+Pb collisions at small $b$ and grows faster
with $b$ due to its smaller system size. 

The minimum transverse momentum cutoff $p_0$ for light flavor minijet
productions is another  key parameter in the HIJING model and thus in
the initial condition of the AMPT
model~\cite{Wang:1991hta,Deng:2010mv,Zhang:2019utb}. 
In our update of the AMPT model with modern
nPDFs~\cite{Zhang:2019utb}, the collision energy dependence of $p_0$
is determined from fitting the $pp$ cross section data. 
Then motivated by the physics of the color glass
condensate~\cite{McLerran:1993ni}, a global nuclear scaling of the 
$p_0$ cutoff~\cite{Zhang:2019utb} has been introduced for central $AA$
collisions above the top RHIC energy of $200A$ GeV to describe the
experimental data on charged particle yields in central Pb+Pb
collisions at LHC energies.  Here~\cite{Zhang:2021vvp} we go beyond the global nuclear
scaling and instead consider $p_0$ as a local variable that depends on
the transverse position of the corresponding hard process in each
event. As $p_0$ is expected to increase with the system size, we
related its value to the nuclear thickness functions in a general $AB$
collision as~\cite{Zhang:2021vvp}
\begin{eqnarray}
\label{eq:p0}
p_0(s_A,s_B,s)&=&p_0^{pp}(s) \left [ \sqrt {T_A(s_A) T_B(s_B)}/T_p \right ]^{\alpha(s)}.
\end{eqnarray}
As $T_A(s) \propto A^{1/3}$, Eq.\eqref{eq:p0} approximately gives
$p_0 \propto A^{\alpha(s)/3}$ for central $AA$ collisions  and thus essentially
recovers the global nuclear scaling if $\alpha(s)=3q(s)$. On
the other hand, for peripheral collisions where $T_A(s_A)$ and
$T_B(s_B)$ are very small and thus replaced with the proton value
($T_p$), Eq.\eqref{eq:p0} automatically gives the $p_0$ value for $pp$
collisions. 

Since $p_0^{pp}(s)$ works for charged particle yields in central
Au+Au collisions at and below $200A$ GeV, we assume that the need to
modify $p_0$ in nuclear collisions starts at the top RHIC
energy~\cite{Zhang:2019utb}. 
From the comparison to charged particle yields in the most central 
Pb+Pb collisions at $2.76A$ and $5.02A$ TeV, we obtain the 
preferred $\alpha(s)$ values at those two energies. 
We then fit the $\alpha(s)$ function as~\cite{Zhang:2021vvp} 
\begin{eqnarray}
\alpha(s)&=&0.0918\ln \left (\frac{\sqrt{s}}{E_0} \right )- 0.00602
\ln^{2}  \left ( \frac{\sqrt{s}}{E_0} \right ) \nonumber \\
&&+0.000134 \ln^{3}  \left ( \frac{\sqrt{s}}{E_0} \right ), {\rm
   ~for~} \sqrt{s} \geq E_0,
\end{eqnarray}
with $\alpha(s)=0$ for $\sqrt{s} < E_0=200$ GeV. We see in
Fig.~\ref{fig:blp0ab}(a) that $\alpha(s) \approx 3q(s)$ as expected. 
We also see that both approach the value of 1/2 at very high energies; 
this is consistent with our expectation~\cite{Zhang:2019utb} that 
$p_0$ is closely related to the saturation momentum $Q_s$ 
in the color glass condensate~\cite{McLerran:1993ni},
where $Q_s \propto A^{1/6}$ in the saturation regime.

Figure~\ref{fig:blp0ab}(b) also shows the average $p_0$ value 
as a function of the impact parameter for Pb+Pb collisions at $2.76A$
TeV and $5.02A$ TeV as well as $p$Pb collisions at $5.02A$ TeV. 
As expected, we see that $\langle p_0 \rangle$ decreases with the
impact parameter and that $\langle p_0 \rangle$ at the lower LHC
energy is smaller than that at the higher LHC energy  due to the
smaller $\alpha(s)$ value. 
Also, $\langle p_0 \rangle$ in $p$Pb collisions is smaller than that in
Pb+Pb collisions at the same colliding energy due to its smaller size. 
In addition, the relative variation of $\langle p_0 \rangle$ with the
impact parameter is seen to be much weaker than that of $\langle b_L
\rangle$ since $\alpha(s) \ll \beta(s)$ for the exponents in the
local nuclear scaling relations. 

\begin{figure}[!htb]
\includegraphics[scale=0.43]{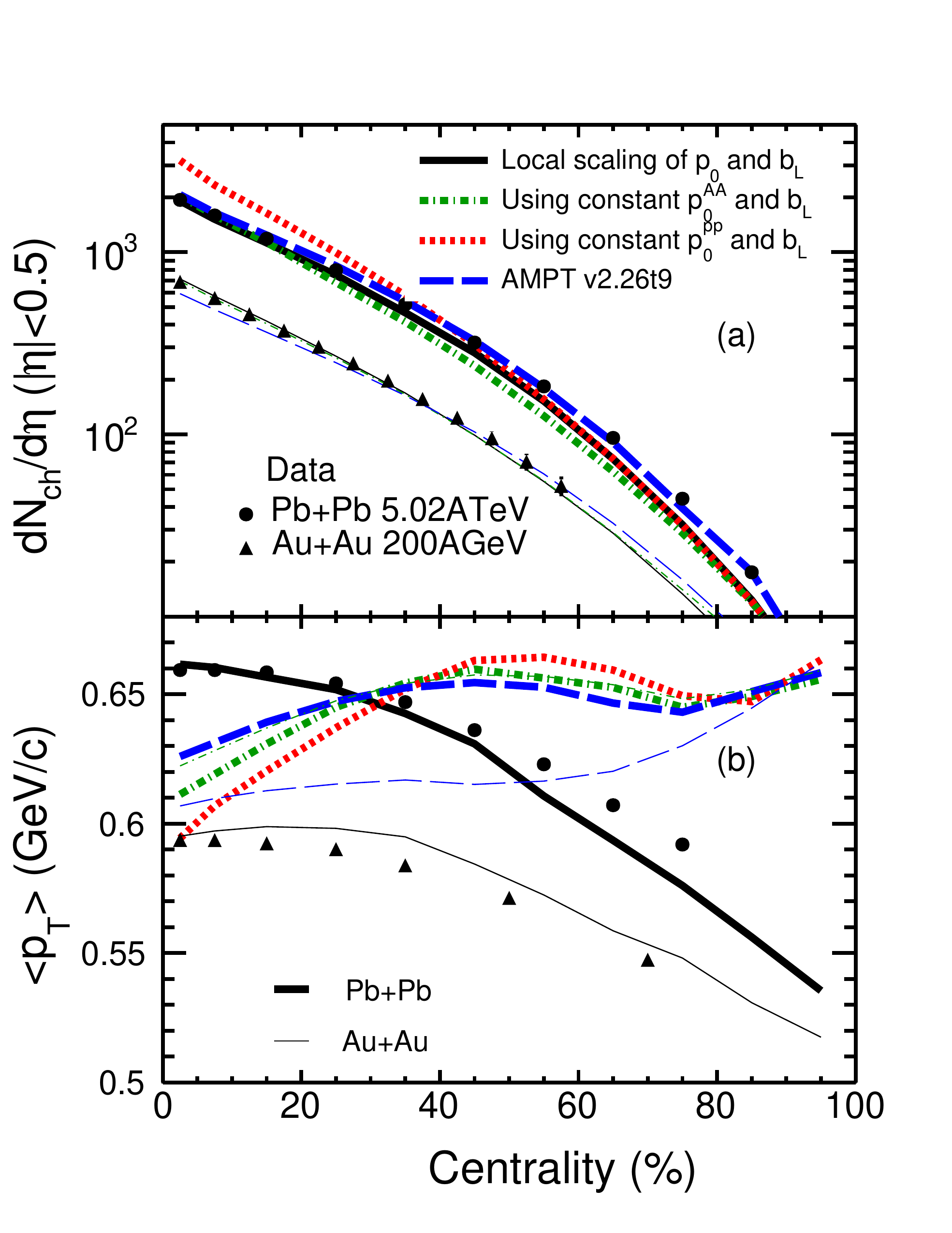}
\caption{(a) $dN_{\rm ch}/d\eta$ and (b) $\langle \pt \rangle$
  around mid-pseudorapidity versus the centrality of $5.02A$ TeV Pb+Pb
  collisions (thick curves) and $200A$ GeV Au+Au collisions (thin
  curves) from this work (solid curves) and earlier AMPT versions in
  comparison with the experimental data (symbols). The   $\pt$ range
  used for the $\langle \pt \rangle$   calculation is [0.15, 2]
  GeV$/c$ at $5.02A$ TeV and [0.2, 2] GeV$/c$ at $200A$ GeV.} 
\label{fig:aa}
\end{figure}

We show in Fig.~\ref{fig:aa} the $dN_{\rm ch}/d\eta$ 
yield in panel (a) and charged particle $\langle \pt \rangle$ in panel
(b) around mid-pseudorapidity versus centrality from different AMPT
versions in comparison with the experimental data for Au+Au collisions
at $200A$ GeV and Pb+Pb collisions at $5.02A$ 
TeV~\cite{Adamczyk:2017iwn,Back:2004ra,Adams:2003kv,Abelev:2012hxa,Acharya:2018qsh,Adare:2015bua,Abelev:2012hxa,Adam:2016ddh}. 
Using the local nuclear scaling, the improved AMPT model (solid
curves) reasonably describes these centrality dependence data in $AA$
collisions at both RHIC and the LHC energies, with a significant 
improvement in the $\langle \pt \rangle$ description
as shown in Fig.~\ref{fig:aa}(b). When we switch
off the local nuclear scaling of $p_0$ and $b_L$ but instead use
constant $b_L=0.15$ GeV$^{-2}$ and $p_0(s)$ (constant at a given
energy),  we recover the AMPT model developed
earlier~\cite{Zheng:2019alz}  and obtain the dot-dashed curves when
using $p_0(s)=p_0^{AA}(s)$ and the dotted curves when using
$p_0(s)=p_0^{pp}(s)$. They both give the wrong centrality dependence
of $\langle \pt \rangle$, since the model results (dot-dashed or
dotted) show a mostly increasing trend from central to peripheral
collisions while the data show a mostly decreasing trend.  
Results from the public AMPT version 2.26t9~\cite{ampt} are also shown (dashed
curves)~\cite{Lin:2014tya}, which also fail to 
describe the centrality dependence of charged particle $\langle \pt 
\rangle$ data. 

In Fig.~\ref{fig:aa}(a), we see that the charged particle yield in
central Pb+Pb collisions at $5.02A$ TeV is significantly overestimated
when using $p_0(s)=p_0^{pp}(s)$, where the global nuclear scaling
$p_0(s)=p_0^{AA}(s)$ is needed to reproduce the particle yield. 
From the Pb+Pb results, we also see that the effect from the global
nuclear scaling of $p_0$ in peripheral collisions is much smaller than
that in central collisions,  because the binary scaling of minijet
productions  makes $p_0$ less important for peripheral collisions. It
is thus not surprising to see that the $dN_{\rm   ch}/d\eta$  results
from the local nuclear scaling are similar to the AMPT results using
the constant $p_0^{AA}$ for central collisions but close to the AMPT
results using the constant $p_0^{pp}$ for peripheral collisions.

\begin{figure}[!htb]
\includegraphics[scale=0.43]{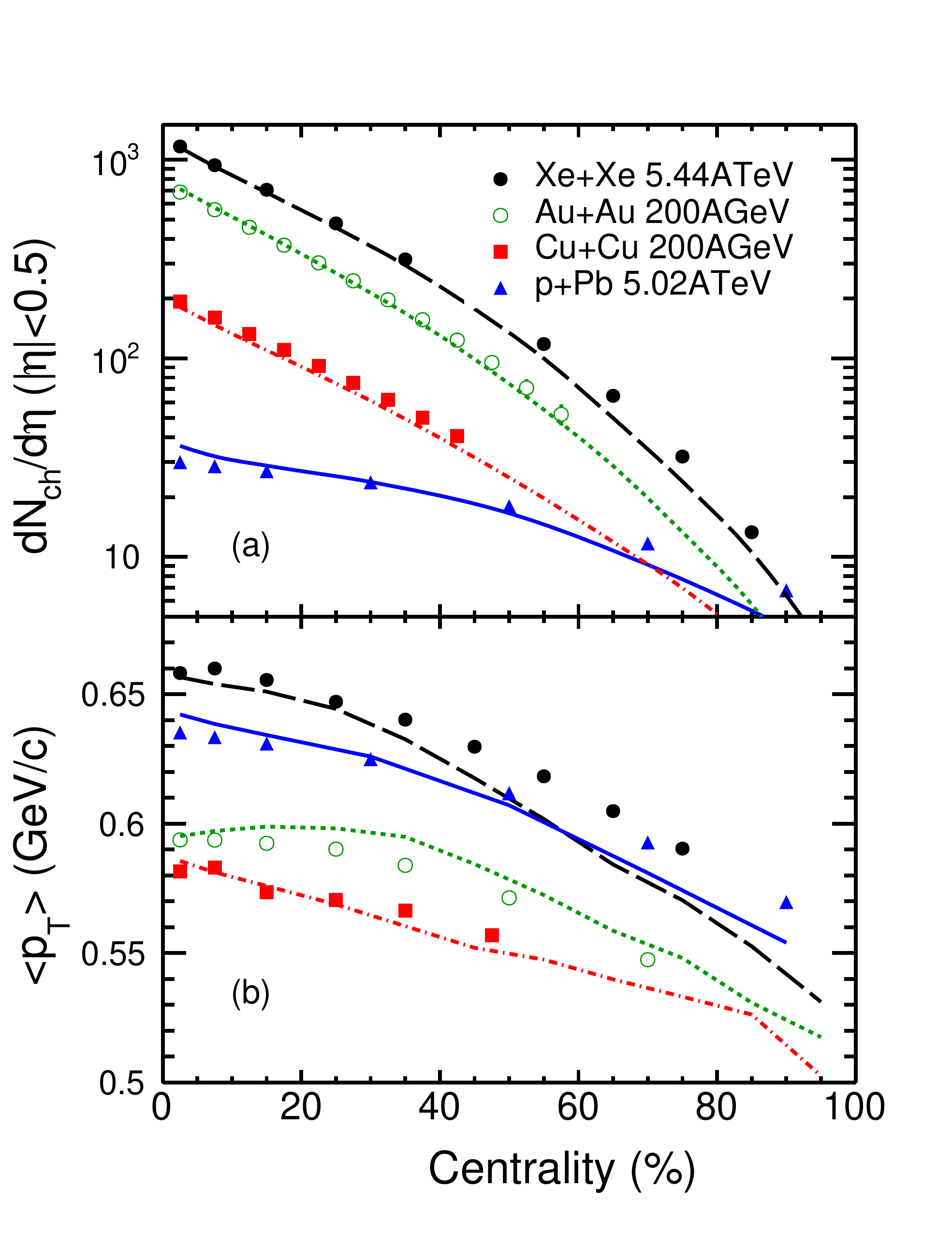}
\caption{(a) $dN_{\rm ch}/d\eta$ and (b) $\langle \pt \rangle$
  around mid-pseudorapidity versus the centrality of Xe+Xe collisions
  at   $5.44A$ TeV, Cu+Cu collisions at $200A$ GeV, and $p$Pb
  collisions at   $5.02A$ TeV from the AMPT model (curves) in
  comparison with the   experimental data (symbols). The $\pt$ range
  used for the   $\langle \pt \rangle$ calculation is [0.15, 2]
  GeV$/c$ at LHC energies and [0.2, 2] GeV$/c$ at $200A$ GeV.}
\label{fig:small}
\end{figure}

The local nuclear scaling relations also predict how observables
depend on the system size going from large to small systems.
Figures~\ref{fig:small}(a) and \ref{fig:small}(b) show respectively the
$dN_{\rm ch}/d\eta$ and charged particle $\langle \pt \rangle$ around
mid-pseudorapidity from the AMPT model~\cite{Zhang:2021vvp}
 versus centrality in comparison with the experimental data for  Au+Au
 collisions and several smaller collision
systems~\cite{Acharya:2018eaq,Acharya:2018hhy,Alver:2005nb,Adare:2015bua,Adam:2014qja,Acharya:2018hhy,Adare:2015bua}. 
We see that the improved AMPT model describes these data rather well,
further demonstrating the validity of the local nuclear scaling
assumption. Note that, although the mid-pseudorapidity $dN_{\rm
  ch}/d\eta$ and $\langle \pt \rangle$ data for the most central
Au+Au/Pb+Pb collisions have been used in the determination of the
parameter functions $\alpha(s)$ and $\beta(s)$, the data of these
smaller systems are not considered in the fitting of the parameters. 
In Fig.~\ref{fig:small} we also see that the changes of the charge
particle yield and $\langle \pt \rangle$ from Cu+Cu to Au+Au
collisions at $200A$ GeV  are well accounted for by the local nuclear
scaling. For example, the $\langle \pt \rangle$ in Cu+Cu is generally
smaller than that in Au+Au due to the larger $b_L$ value for Cu+Cu 
collisions. Note however that our calculations here have not considered the
deformation of the Xe nucleus~\cite{Moller:2015fba}. 

\subsection{PYTHIA8 initial condition with sub-nucleon structure}
\label{subsec:pythia8}

The modifications of the AMPT initial condition discussed so far have
been performed within the framework of the HIJING two-component
model that uses the PYTHIA5 program. While the development of local
nuclear scaling~\cite{Zhang:2021vvp} enables the AMPT model to
reproduce the system size dependence and centrality dependence of
changed particle yields and $\langle \pt \rangle$ in $p$A and $AA$
collisions using the parameter values for minimum bias $pp$
collisions, we have not directly addressed the multiplicity dependence
of these observables, especially the $\langle \pt \rangle$, in $pp$
collisions.  On the other hand, PYTHIA8~\cite{Sjostrand:2014zea} is
quite successful in describing the particle production in $pp$
collisions. It has been extended to treat $p$A or $AA$
collisions based on the Angantyr framework~\cite{Bierlich:2018xfw},
and PYTHIA8 has been used as the initial condition generator for
multiple heavy ion Monte Carlo
models~\cite{Kanakubo:2019ogh,Papp:2018qrc,Putschke:2019yrg}.  
Therefore, it is worthwhile to have the option to use PYTHIA8 as the 
initial condition for the AMPT model.  

Recently we have coupled PYTHIA8 with the final state parton and hadron
interactions and quark coalescence~\cite{He:2017tla} of the AMPT-SM 
model to study $pp$ collisions~\cite{Zheng:2021jrr}. In this approach,
the fluctuating initial condition of AMPT 
originally provided by the HIJING model is replaced by the
PYTHIA/Angantyr model~\cite{Bierlich:2018xfw}. 
In addition, the sub-nucleon structure, which could be important for
collectivity observables in small
systems~\cite{Schenke:2014zha,Mantysaari:2016ykx,Welsh:2016siu,Mantysaari:2017cni,dEnterria:2010xip}, can be modeled when implementing the space-time
structure of the string system generated by PYTHIA. With the proton
charge distribution given by 
\begin{equation}
\rho(r)=\frac{1}{8\pi R^3}e^{-r/R}
\end{equation}
with $R=0.2$ fm, the sub-nucleon spatial structure can be related
to the transverse positions of the excited strings in two ways. In the
first way, the transverse coordinates of the produced string objects
are sampled according to the overlap function of a $pp$ collision
at a given impact parameter $b$:
\begin{equation}
T(x, y, b)=\int\rho(x-b/2, y, z) \rho(x+b/2, y, z) dz,
\end{equation}
where $z$ is along the beam directions. In the second way, the initial
transverse spatial condition including event-by-event sub-nucleon
fluctuations is generated with a Glauber Monte Carlo method based on
the constituent quark
picture~\cite{Eremin:2003qn,Zheng:2016nxx,Loizides:2016djv,Bozek:2016kpf,Welsh:2016siu,Bozek:2019wyr}. By
modeling the proton as three constituent quarks, the interaction of
two protons can be interpreted as collisions between the constituent 
quarks from each incoming proton within the Glauber model
framework~\cite{Loizides:2016djv,dEnterria:2020dwq}. The 
positions of the quark constituents are first sampled with the proton
profile $\rho(r)$, then the transverse coordinates of the excited 
strings are randomly assigned to the binary collision center of each
interacting constituent pair. 

\begin{figure}
\centering
\includegraphics[width=0.49\textwidth]{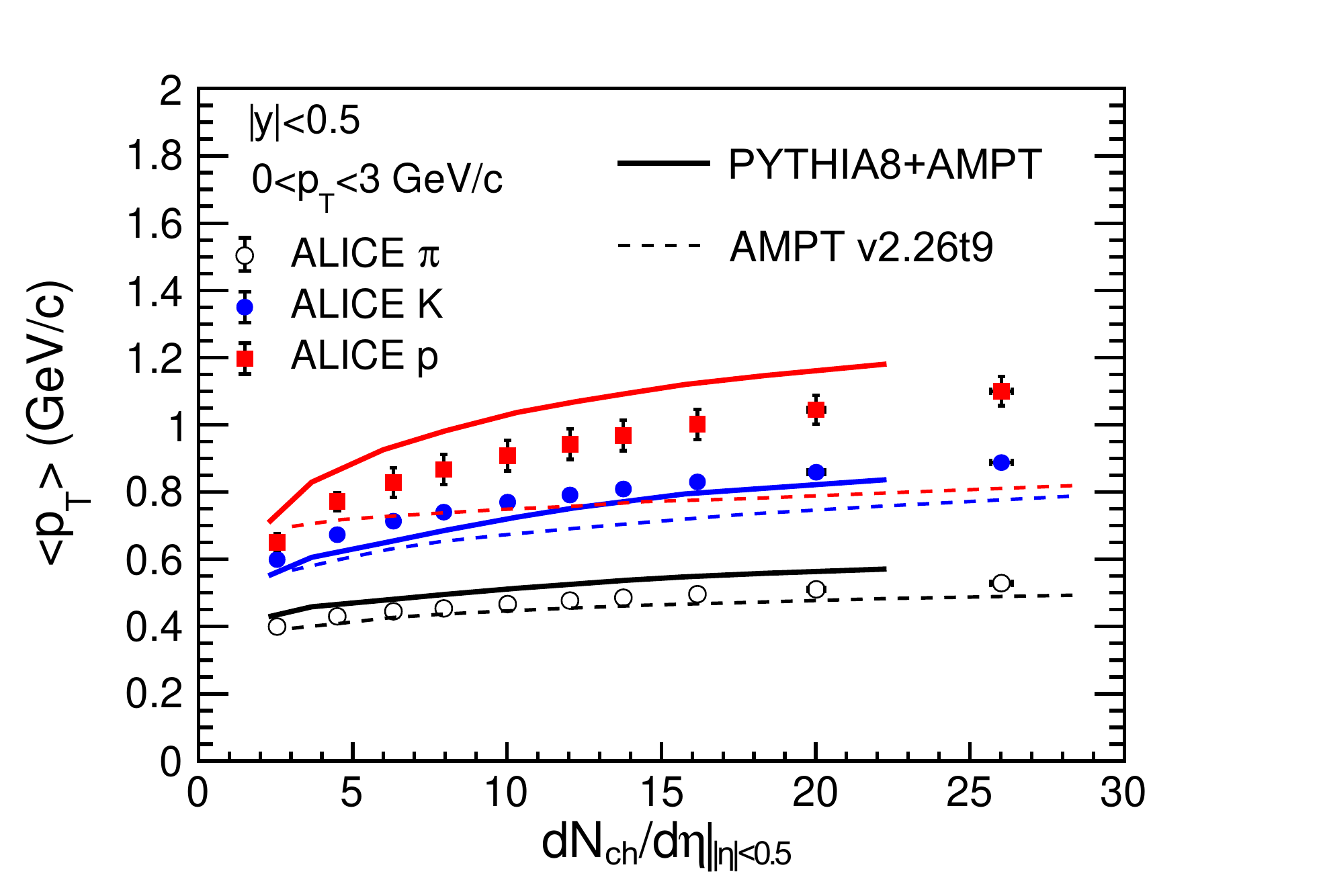}
\caption{$\langle \pt \rangle$ of $\pi$ (black), $K$ (blue) and
  proton (red) at mid-rapidity within $0<\pt<3$ GeV$/c$ versus the
  charged hadron multiplicity  density in 13 TeV $pp$ collisions. The 
  AMPT model using the PYTHIA8 initial condition (solid curves)
  are compared to the original AMPT model (dashed curves) and the
  ALICE data.}
\label{fig:Avgpt_pikp}
\end{figure}

Figure~\ref{fig:Avgpt_pikp} shows the effect of using PYTHIA8
as the AMPT initial condition on the identified particle
$\langle \pt \rangle$ versus  the charge particle pseudo-rapidity
density in $pp$ collisions at $\sqrt{s}=13$ TeV.
Note that only hadrons within $0<\pt<3$ GeV$/c$ and $|y|<0.5$ are
included in this comparison, and the central values of ALICE data are
obtained with a refit to the data~\cite{Acharya:2020zji}. 
We see that this AMPT model (solid curves), which uses the PYTHIA8
initial condition and includes both parton and hadron evolutions,
roughly reproduces the experimental data.  
On the other hand, the original AMPT model (dashed curves) reasonably
describes the pion $\langle \pt \rangle$ but gives a very weak
multiplicity dependence for the proton $\langle \pt \rangle$. The
significant improvement compared to the original AMPT model on the
multiplicity dependence of the proton $\langle \pt \rangle$ presumably
results from multiparton interaction in the PYTHIA8 model. 
 
\begin{figure*}[hbt!]
\centering
\includegraphics[width=0.6\textwidth]{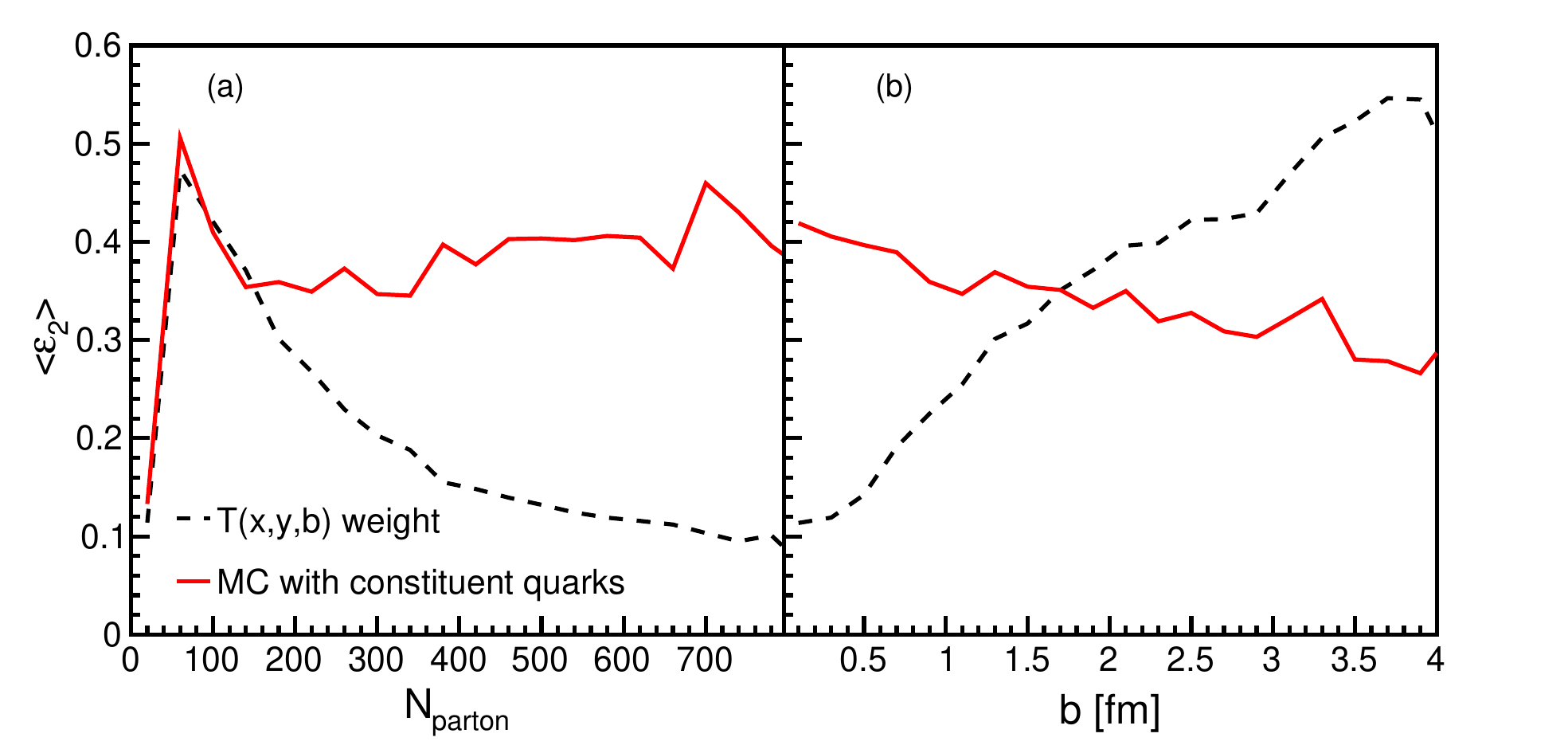}
\includegraphics[width=0.314\textwidth]{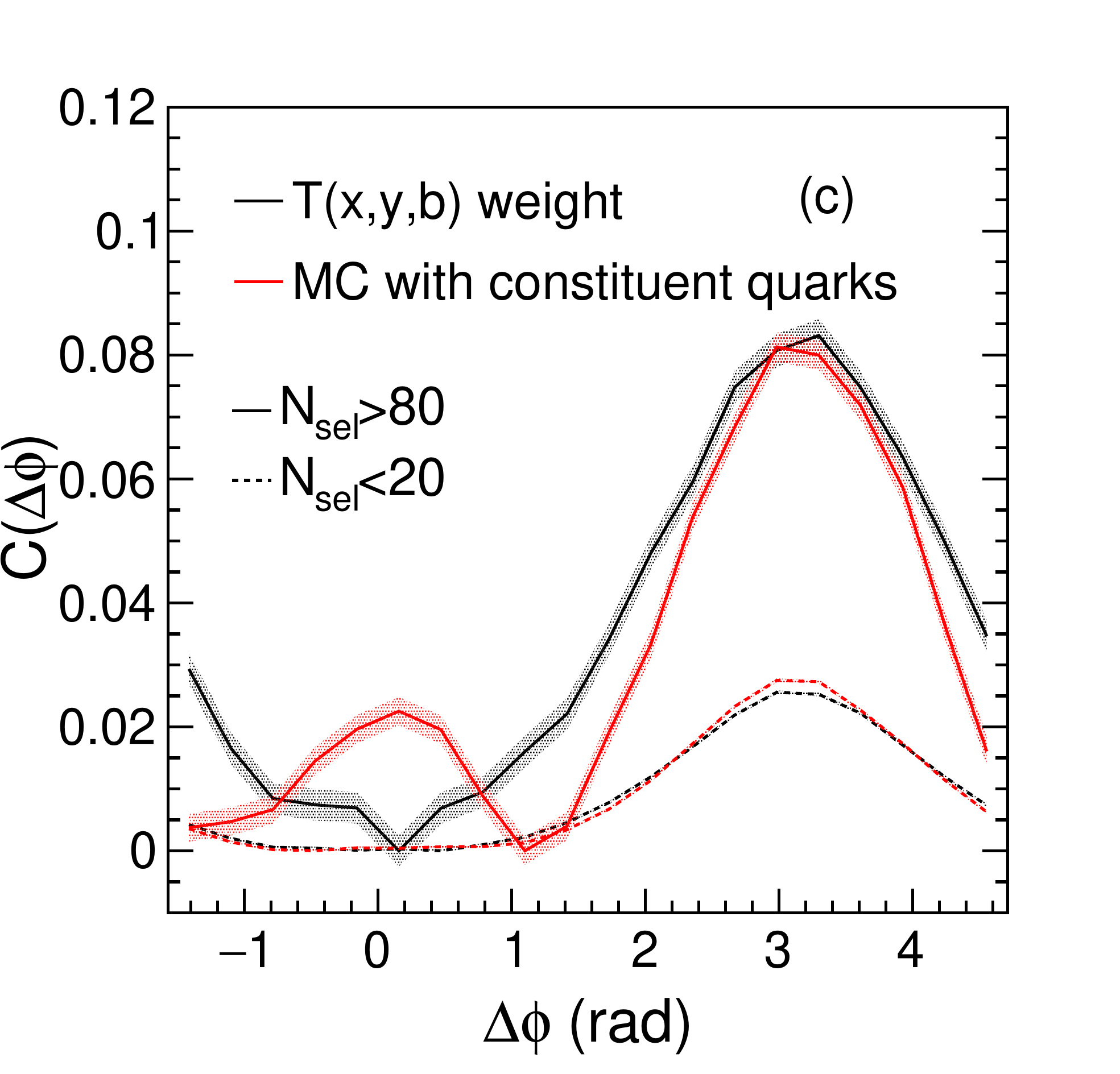}
\caption{The initial eccentricity of partons right after string
  melting versus (a) the number of partons and (b) the impact
  parameter, and (c) two-particle long-range angular correlations for
  events at two different multiplicity classes, for $pp$ collisions at 
  13 TeV. AMPT results with the sub-nucleon structure are shown
  for the overlap function method (black curves) and the MC method
  with constituent quarks (red curves).} 
\label{fig:ecc2}    
\end{figure*}

Figure.~\ref{fig:ecc2}(a) shows the average initial spatial
eccentricity of partons in the transverse plane right after string
melting as a function of the parton multiplicity of each event 
from the two ways of generating the sub-nucleon spatial structure.  
Note that only partons with formation time less than 5 fm/c are
considered, and eccentricities are calculated with the initial
position of each parton at its formation
time~\cite{Nagle:2017sjv}. When using the overlap function weighting
method (black curves), the eccentricity is largely driven by the
geometric shape of the transverse overlap area and thus decreases
significantly with the parton multiplicity as shown in panel (a) and
increases significantly with the impact parameter as shown in panel
(b).  On the other hand, when using the Monte Carlo method with 
constituent quarks (red curves), large eccentricities in the initial
condition can be generated even in very central collisions or 
events at high multiplicities. Figure~\ref{fig:ecc2}(b) actually shows
that the initial eccentricity from the constituent quark method is
larger for $pp$ collisions at smaller impact parameters, opposite to
the behavior from the overlap function method. 

The difference in the initial spatial eccentricity could certainly
affect final state momentum anisotropies in small collision systems
after interactions in the AMPT model convert the spatial anisotropies
into momentum anisotropies~\cite{Lin:2001zk,He:2015hfa,Lin:2015ucn}. 
Using the AMPT model with PYTHIA8 as the initial condition, we have 
found~\cite{Zheng:2021jrr} that two-particle long-range correlations
in high multiplicity $pp$ collisions at the LHC  depend sensitively on
how the sub-nucleon structure of the proton is implemented. 
We analyze the projected correlation function of two charged
hadrons with a large pseudorapidity gap:
\begin{equation}
C(\Delta\phi)=\frac{1}{N_{trig}}\frac{dN^{pair}}{d\Delta\phi}.
\end{equation}
Both trigger and associate hadrons are required to be within $1<\pt<3$
GeV/$c$ and $|\eta|<2.4$ following the analysis procedure of the CMS
Collaboration~\cite{Khachatryan:2016txc}, and the two hadrons in each
pair must be separated in pseudo-rapidity with a gap $|\Delta\eta|>2$.
Events are separated into two categories based on $N_{sel}$, the
number of selected charge tracks with $\pt>0.4$ GeV/$c$ and
$|\eta|<2.4$. High multiplicity events are defined as those with
$N_{sel}>80$, while low multiplicity events are defined as those with
$N_{sel}<20$.

Figure~\ref{fig:ecc2}(c) shows the multiplicity dependence of the
$C(\Delta\phi)$ function from the two ways of generating the
sub-nucleon spatial structure for 0.2 mb parton cross section
~\cite{Zheng:2021jrr}.  We see that the AMPT model using
PYTHIA8 shows a long-range ridge-like 
structure for high multiplicity events when the proton geometry is
modeled with the constituent quark method (red solid curve), while
the overlap function weighting method (black solid curve) does not
show this structure. This demonstrates the connection between
two-particle long-range correlations and the underlying sub-nucleon
structure and fluctuations. Note that a significant near-side ridge
structure in the correlation function is found in the experimental
data, which has been regarded as an important signature of
collectivity in high multiplicity $pp$
events~\cite{Khachatryan:2010gv,Khachatryan:2016txc}. 

We note that the original AMPT-SM model also shows the long-range
near-side correlations, although it does not include the sub-nucleon
structure~\cite{Zheng:2021jrr}. 
In addition, the PYTHIA event generator itself has considered final
state hadronic
rescatterings~\cite{Sjostrand:2020gyg,Ferreres-Sole:2018vgo,daSilva:2020cyn,Bierlich:2021poz}.
Using the AMPT-SM model with PYTHIA8 initial conditions, we can extend
the study of $pp$ collisions~\cite{Zheng:2021jrr}  to p$A$ and $AA$ 
collisions with the Angantyr model within the PYTHIA8 framework. That
would lay a solid foundation for the studies of different mechanisms
of collectivity,  such as string shoving and parton/hadron evolutions,
with the same model. 

\subsection{Improved algorithm for the parton cascade}
\label{subsec:zpc}

Particle correlations and momentum anisotropies in the AMPT-SM model 
are usually dominated by parton
interactions~\cite{Lin:2001zk,Lin:2004en,Li:2016flp}. 
We have also found that even a few parton scatterings in a small
system is enough to generate significant  momentum
anisotropies through the parton escape
mechanism~\cite{He:2015hfa,Lin:2015ucn}. 
It is therefore important to ensure that the parton cascade solution
in the AMPT model is accurate. 
 
The ZPC elastic parton cascade~\cite{Zhang:1997ej} in the AMPT model
solves the Boltzmann equation by the cascade method, where a scattering
happens when the closest distance between two partons is less than the
range of interaction $\sqrt{\sigma_p/\pi}$ with $\sigma_p$ being the
parton scattering cross section. The default differential cross
section in ZPC for two-parton scatterings, based on the gluon elastic
scattering cross section as calculated with QCD at leading order, is
given by~\cite{Zhang:1997ej,Lin:2004en} 
\begin{equation}
\frac{d\sigma_p}{d \hat t}=\frac{9\pi \alpha_{s}^{2}}{2} \left
  (1+\frac{\mu^2}{\hat s} \right ) \frac{1}{(\hat t-\mu ^{2})^{2}},
\label{eq:dsigmadt}
\end{equation}
where $\mu$ is a screening mass to regular the total cross
section. This way the total cross section has no explicit dependence
on $\hat s$: 
\begin{equation}
\sigma_p = \frac{9\pi \alpha_{s}^{2}}{2\mu ^{2}}.
\label{eq:sigma}
\end{equation}
The above Eqs.(\ref{eq:dsigmadt}-\ref{eq:sigma}) represent
forward-angle scatterings. For isotropic scatterings, $d\sigma_p/d\hat
t$ is independent of the scattering angle.

It is well known that cascade calculations suffer from the causality 
violation~\cite{Kodama:1983yk,Kortemeyer:1995di} due to the
geometrical interpretation of cross section. 
This leads to inaccurate numerical results at high densities and/or
large scattering cross sections (i.e., large opacities).  
For example, a recent study~\cite{Molnar:2019yam} has shown that the
effect of causality violation on the elliptic flow from the AMPT-SM
model~\cite{Lin:2004en} is small but non-zero. Causality violation
also leads to the fact that different choices of performing collisions
and/or the reference frame can lead to different numerical
results~\cite{Zhang:1996gb,Zhang:1998tj,Cheng:2001dz}.  
These numerical artifacts due to the causality violation can be
reduced or removed by the  parton subdivision method
~\cite{Wong:1982zzb,Welke:1989dr,Kortemeyer:1995di,Pang:1997,Zhang:1998tj,Molnar:2000jh,Molnar:2001ux,Molnar:2004yh,Xu:2004mz}.  
However, parton subdivision usually alters the event-by-event
correlations and fluctuations, the importance of which has been more
appreciated in  recent years~\cite{Alver:2010gr}; it is also much more
computationally expensive. 
Therefore, it is preferred to improve the parton cascade to yield
solutions that are accurate enough without using parton
subdivision. We have recently pursued this goal for box
calculations~\cite{Zhao:2020yvf}.  

In ZPC, one can take different choices or collision schemes to
implement the cascade method~\cite{Zhang:1997ej}. 
With the closest approach criterion for parton scatterings, the
closest approach distance is usually calculated in the two-parton
center of mass frame. 
Two partons may collide when their closest approach distance is
smaller than $\sqrt{\sigma_p/\pi}$, and at a given global time all such
possible collisions in the future are ordered in a collision list with
the ordering time of each collision so that 
they can be carried out sequentially. The collision list is updated
continuously after each collision, 
and for expansion cases the parton system dynamically freezes out when
the collision list is empty. 
For calculations of a parton system in a box, we terminate the parton
cascade at a global time that is large enough so that the parton
momentum distribution changes little afterwards. 
When the closest approach distance is calculated in the two-parton
center of mass frame, the collision time of a scattering 
is a well-defined single value.  
However, because of the finite $\sigma_p$ the two partons have
different spatial coordinates in general; this collision time in the
two-parton center of mass frame thus becomes two different collision 
times in the global frame (named here as $ct_1$ and $ct_2$
respectively for the two colliding partons) after the Lorentz
transformation. The default collision scheme of
ZPC~\cite{Zhang:1997ej}  uses $(ct_1+ct_2)/2$ as both the collision
time and ordering time; this is the case for the AMPT
model~\cite{Lin:2004en} 

Results from the default ZPC scheme~\cite{Zhao:2020yvf} at
$\sigma_p=2.6$ mb  are shown in Fig.~\ref{fig:scheme} (curves with
open circles). Panel (a) shows the final parton $\pt$
distribution, while panels (b) and (c) show the time evolution of
parton $\langle \pt \rangle$ (scaled by $T$) and variance of $\pt$
(scaled by $T^2$), respectively. 
The gluon system is initialized in a box with an off-equilibrium
initial momentum distribution as shown by the dot-dashed curve in
panel (a), where the gluon density is set the same as that for a
thermalized gluon system with the Boltzmann distribution 
at temperature $T=0.5$ GeV. 
We see from Fig.~\ref{fig:scheme}(a) that the final distribution from
the default ZPC scheme deviates considerably from the expected thermal
distribution (dotted curve). On the other hand, we find that a new
collision scheme, which uses $min(ct_1,ct_2)$ as both the collision
time and ordering time, gives a final distribution very close to the
thermal distribution~\cite{Zhao:2020yvf}.  
The causality violation usually suppresses collision rates, which is
the case for the default ZPC scheme; it is therefore understandable
that choosing time $min(ct_1,ct_2)$ instead of $(ct_1+ct_2)/2$  
enhances the collision rates and thus suppresses the causality violation.  

We use the parton subdivision method to obtain the ``exact'' time
evolutions of $\langle \pt \rangle$ and $\pt$ variance (dashed
curves) in Figs.~\ref{fig:scheme}(b) and (c). 
We see that the time evolution of the $\pt$ variance from the default
scheme deviates significantly from the ``exact'' parton subdivision
result, although the time evolutions of $\langle \pt \rangle$ are close
to each other (mostly due to the conservation of total momentum).
In contrast, the time evolution of the $\pt$ variance from the new scheme
~\cite{Zhao:2020yvf} is very close to the parton subdivision result,
which at late times agrees with theoretical expectation (diamond). 
By examining cases of different parton densities and cross
sections~\cite{Zhao:2020yvf}, we find rather surprisingly that the new
scheme for ZPC gives very accurate results (i.e., very close to parton
subdivision results and/or theoretical values) even at very large
opacities, such as the case of $T=0.7$ GeV and $\sigma_p=10$ mb.

\begin{figure*}[!htb]
\includegraphics [width=0.9\hsize]  {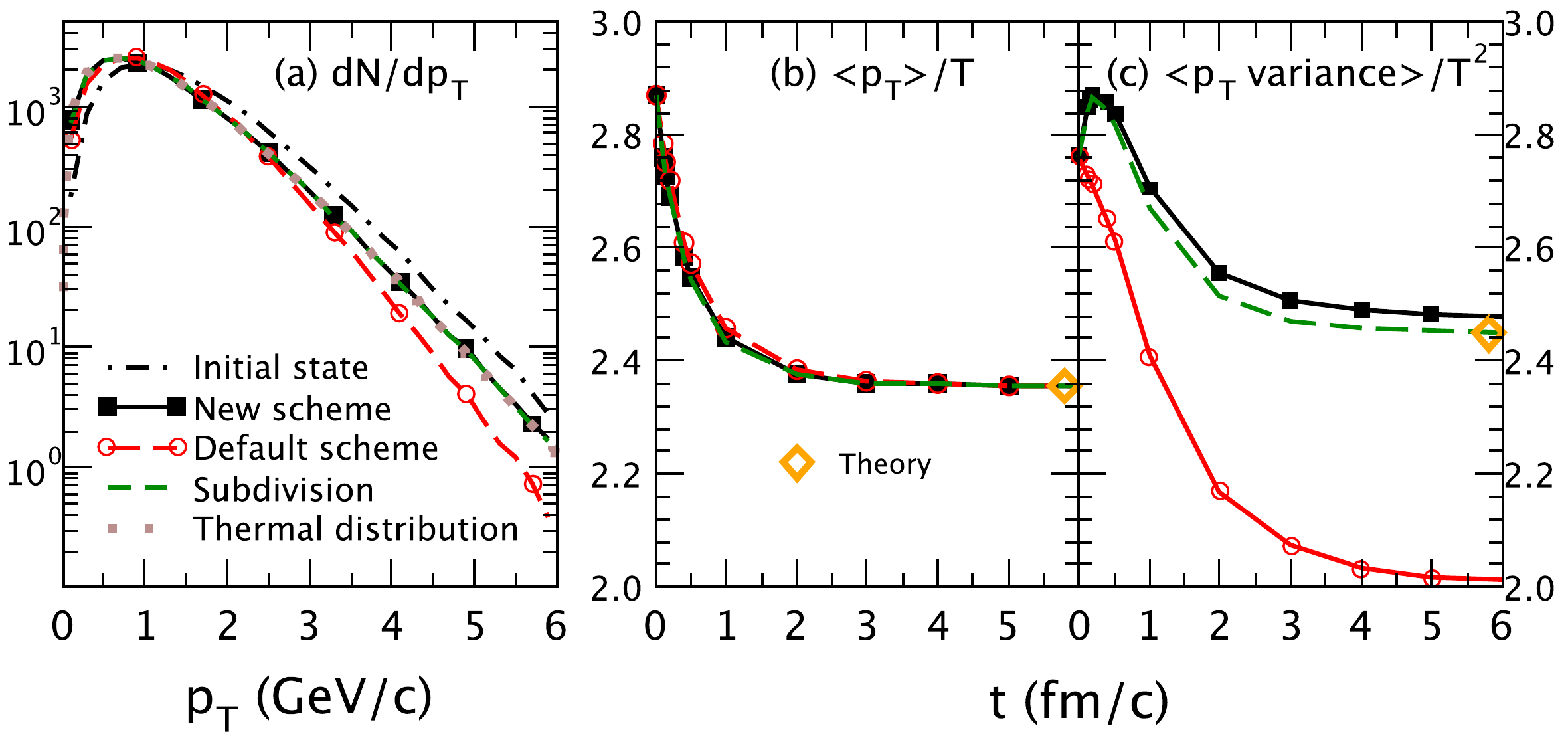}
\caption{(a) The final $\pt$ distribution, (b) time evolution of
  $\langle \pt \rangle/T$, and (c) time evolution of $\langle \pt$
  variance$\rangle/T^2$ from ZPC from the default collision scheme 
  (open circles) and the new collision scheme (filled squares) in
  comparison with parton subdivision results at subdivision factor
  $l=10^6$ (dashed curves) for gluons in a box at $T=0.5$ GeV and
  $\sigma_p=2.6$ mb.}  
\label{fig:scheme}
\end{figure*}

We have used a novel parton subdivision method for the
results shown in Fig.~\ref{fig:scheme}.
In the standard method, one increases the initial parton
number per event by factor $l$ while decreasing the cross section by
the same factor, which can be schematically represented by the
following:
\begin{equation}
N \rightarrow l \times N, ~~V {~\rm unchanged},
\label{eq:subd3}
\end{equation}
where $N$ is the initial parton number in an event and $V$ is the
initial volume of the parton system. 
Since the number of possible collisions scales with $l^2$,
the subdivision method is very expensive in terms of the computation
time, which roughly scales with $l^2$ per subdivision event or 
$l$ per simulated parton. 
However, for box calculations where the density function $f(\bm x,\bm
p,t)$ is spatially homogeneous, the following new subdivision method
can be used: 
\begin{equation}
N {~\rm unchanged}, ~~V \rightarrow V/l, 
\label{eq:subd4}
\end{equation}
where we decrease the volume of the box by factor $l$ while keeping
the same parton number and momentum distribution in each event. 
This subdivision method is much more efficient than the standard
subdivision method; we therefore use a huge subdivision factor
$10^6$ (instead of the usual value of up to a few hundreds). 

We emphasize that the differential cross section must not be
changed when performing parton subdivision; as a result, the exact 
transformation for parton subdivision is~\cite{Zhao:2020yvf} 
\begin{equation}
f(\bm x,\bm p,t) \rightarrow  l \times f(\bm x,\bm p,t),
~~~\frac{d\sigma_p}{d\hat t}  \rightarrow \frac{d\sigma_p}{d\hat t} / l.
\label{eq:subd2}
\end{equation}
This is especially relevant for forward-angle scattering. 
For example, when parton subdivision requires the decrease of the
forward-angle cross section of Eq.(\ref{eq:sigma}), one should not do
that by increasing the screening mass $\mu$ by a factor of $\sqrt{l}$
because that would change the angular distribution of the scatterings
in Eq.(\ref{eq:dsigmadt}).  Instead, one can decrease the $\alpha_s$
parameter by a factor of $\sqrt{l}$, which decreases the total
scattering cross section while keeping its angular distribution the
same. 

\begin{figure}[!htb]
\includegraphics [width=0.9\hsize]  {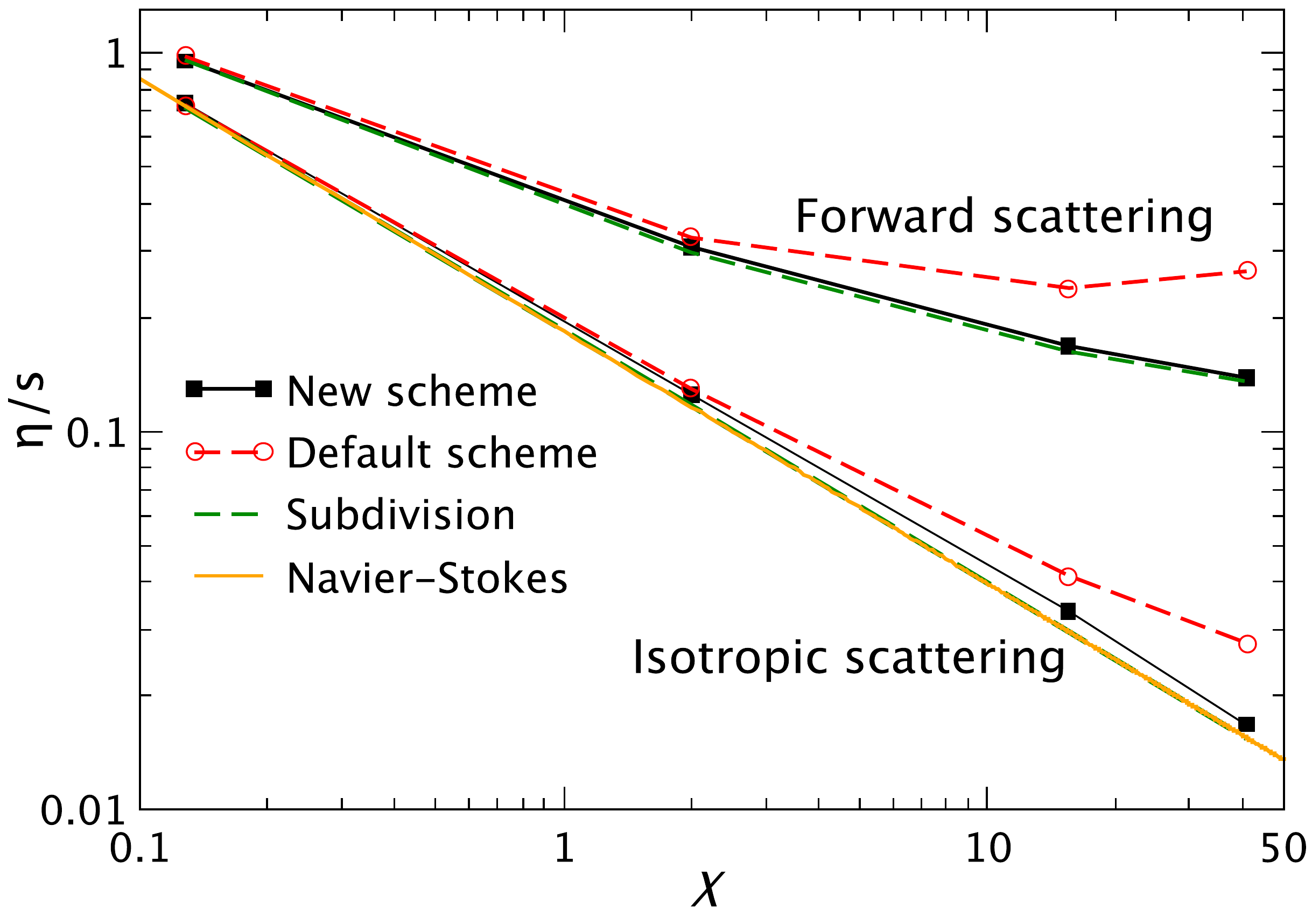}
\caption{The $\eta/s$ ratio for different cases of gluon scatterings
  in a box versus the opacity $\chi$; the solid curve without symbols
  represents the Navier-Stokes expectation for isotropic scatterings.
}
\label{fig:etas}
\end{figure}

Transport coefficients such as the shear viscosity $\eta$ represent 
important properties of the created matter
\cite{Schafer:2009dj}. Therefore, we have also evaluated the effect of
the new collision scheme on the shear viscosity $\eta$ and its ratio
over the entropy density $\eta/s$. 
The Green-Kubo relation~\cite{Green,Kubo} has been 
applied~\cite{Muronga:2003tb,Demir:2008tr,Fuini:2010xz,Wesp:2011yy,Li:2011xu}
to calculate the shear viscosity at or near equilibrium.  
We thus start with an equilibrium initial condition for shear viscosity
calculations according to the Green-Kubo relation~\cite{Muronga:2003tb}. 

Figure~\ref{fig:etas} shows our $\eta/s$ results  
as functions of the opacity parameter
$\chi$, which is defined as~\cite{Zhang:1998tj}
\begin{equation}
\chi =\sqrt{\frac{\sigma_p}{\pi }} /\lambda
=n \sqrt{\frac{\sigma_p^3}{\pi } },
\label{chi1}
\end{equation}
where $n$ is the parton density and $\lambda$ is the mean free path.
The case shown in Fig.~\ref{fig:scheme} for gluons in a box at $T=0.5$
GeV and $\sigma_p=2.6$ mb then corresponds to $\chi=2.0$,  
and other $\chi$ values shown in Fig.~\ref{fig:etas} are obtained for
the following cases: $T=0.2$ GeV and $\sigma_p=2.6$ mb, $T=0.7$ GeV
and $\sigma_p=5.2$ mb,  and $T=0.7$ GeV and $\sigma_p=10$
mb~\cite{Zhao:2020yvf}.  
For isotropic scatterings of a massless Maxwell-Boltzmann
gluon gas in equilibrium (where $s=4n$ and degeneracy factor
$d_g=16$), we have the following Navier-Stokes expectation:
\begin{equation}
\left (\frac{\eta}{s} \right )^{N\!S} 
\simeq \frac{0.4633}{d_g^{1/3}\chi^{2/3}}
= \frac{0.1839}{\chi^{2/3}},
\label{eq:etasNS}
\end{equation}
which only depends on the opacity $\chi$.
We see in Fig.~\ref{fig:etas} that for isotropic scatterings the
subdivision result agrees well with the Navier-Stokes expectation
(solid curve). On the other hand, the extracted $\eta$ and $\eta/s$ 
values from the default ZPC scheme are significantly different from
the Navier-Stokes expectation or the parton subdivision results at
large opacities, although they agree at low opacities as expected. 
We also see that the results from the new collision scheme are very
close to the subdivision results for both forward-angle scatterings
and isotropic scatterings, even at a huge opacity $\chi=41$. 
The new ZPC collision scheme for box calculations is the first step
towards the validation and improvement of the ZPC parton cascade for
scatterings in 3-dimensional expansion cases.

\section{Other developments}
\label{sec:other}

There are other developments of the AMPT model that have not been
covered in the previous section. Here we gave a brief overview of
some of these works. 

The AMPT model has been extended to include deformed nuclei as
the projectile and/or target. First, deformed uranium nuclei are
implemented~\cite{Haque:2011aa} to study various observables in U+U
collisions at $200A$ GeV and the effect of nuclear deformation. 
Later, the AMPT model is modified to specify 
the initial proton and neutron spatial distributions in the $^{96}$Ru 
or $^{96}$Zr nucleus according to the density functional theory (DFT)
calculations~\cite{Xu:2017zcn, Xu:2017qfs, Li:2018oec}. 
The effects of the DFT nuclear density distributions on the
backgrounds and possible signals of the chiral magnetic effect (CME)
in isobar collisions are then investigated~\cite{Xu:2017zcn}. 
The extended AMPT model is also used in the study 
that proposes a novel method to search for the CME in a single
heavy ion collision system~\cite{Xu:2017qfs}. 
Another study~\cite{Li:2018oec} uses the model to study multiplicity
distributions and elliptic flow in isobar collisions, where the
differences between the two isobar systems have the potential to 
decisively discriminate DFT nuclear distributions from the usual
Woods-Saxon density distributions. 

The AMPT model has also been extended to 
include mean field potentials in the hadronic phase in a study of 
the elliptic flow splitting of particles and antiparticles
at the RHIC BES energies~\cite{Xu:2012gf}. 
A later study couples the AMPT model with a parton transport based on
the 3-flavor Nambu-Jona-Lasinio model~\cite{Xu:2013sta} to
include the partonic mean field potentials; it shows that a
combination of partonic and hadronic mean field potentials 
can describe the observed splitting of elliptic flows.

The current AMPT model has been known to violate the electric charge
conservation because of two reasons~\cite{Lin:2014uwa}. 
First, the hadron cascade is based on the ART model~\cite{Li:1995pra}
that has $K^+$ and $K^-$ as explicit particles but not  $K^0$ or
${\bar K^0}$. As a result, we change $K^0$ to $K^+$ and change $\bar
K^0$ to $K^-$ prior to the hadron cascade in order
to include hadronic interactions of neutral kaons, and after the
hadron cascade we assume the isospin symmetry and thus change half of
the final $K^+$ into $K^0$ and change half of the final $K^-$ to $\bar
K^0$.   The second reason is that many hadron reactions and some
resonance decays in AMPT  violate the electric charge
conservation. Some reaction channels do not consider electric charges
of the initial-state hadrons; instead the isospin-averaged cross
section is used and the electric charge of each final state hadron is 
set randomly~\cite{Lin:2014uwa}. 
We have developed a version of the AMPT model that has corrected 
these problems and thus satisfies the electric charge
conservation~\cite{charge}. This charge-conserved version of the AMPT
model has been  shared with some colleagues for their recent studies
of charge-dependent CME
signals~\cite{Tang:2019pbl,Choudhury:2019ctw}. 

Recently we have developed a pure hadron cascade version of the AMPT
model (AMPT-HC)~\cite{Yong:2021npa} to study heavy ion collisions at
low energies below a few GeVs. 
Note that the Eikonal formalism, which is a basis of the  HIJING model
and thus the initial condition of the standard AMPT model,  is
expected to break down for nuclear collisions at low enough energies.  
We thus treat a heavy ion collision as individual nucleon-nucleon
collisions in the AMPT-HC model.
First, we use the Woods-Saxon nucleon density distribution and the local
Thomas-Fermi approximation to initialize the position and momentum of
each nucleon in the incoming nuclei. 
Primary nucleon-nucleon collisions are then treated with the hadron
cascade component of AMPT, without going through the Lund string
fragmentation, the parton cascade, or quark coalescence.   
In addition to the usual elastic and inelastic collisions, the hadron
cascade in the AMPT-HC model also includes hadron mean field
potentials for kaons, baryons and antibaryons. 
This model has been used to study the $\Xi^{-}$ production in
low energy Au+Au collisions, which is proposed as a better 
probe of the nuclear equation of state at high densities than 
single strangeness (kaon or $\Lambda$)
productions~\cite{Yong:2021npa}. 

\section{Summary and outlook}
\label{sec:summary}

A multi-phase transport model was constructed to provide a
self-contained kinetic theory-based description of 
relativistic nuclear collisions with its four main
components: the fluctuating initial condition, partonic interactions,
hadronization, and hadronic interactions. 
Here we review the main developments since the public release of the
AMPT source code in 2004 and the 2005 publication that described the
details of the model at that time. 
Several developments have been carried out to improve the initial condition, 
including the incorporation of finite nuclear thickness
relevant for heavy ion collisions below the energy of tens of GeVs, the
incorporation of modern parton  distribution functions of nuclei for
high energy heavy ion collisions, improvement of heavy quark
productions, the use of local nuclear scaling of key input parameters
for the system size dependence and centrality dependence, and the
incorporation of PYTHIA8 and sub-nucleon structure. 
There are also ongoing efforts to improve the accuracy of 
the parton cascade without using the parton subdivision method  
that would alter event-by-event correlations and fluctuations. 
In addition, the spatial quark coalescence model has been further
developed to allow a quark the freedom to form either a meson or a
baryon depending on the distance to its coalescing partner(s), which 
improves baryon and antibaryon productions of the model. 
Furthermore, deuteron production and annihilation processes have been
included in the hadron cascade, an AMPT version that satisfies the
electric charge conservation has been developed, and a pure hadron
cascade version of the AMPT model is recently developed to study heavy
ion collisions at low energies below a few GeVs.  
For high energy nuclear collisions where the quark-gluon plasma is
expected, the string melting version of the AMPT model can now
reasonably and simultaneously describe the yield, transverse momentum
spectrum and elliptic flow of the bulk matter from small to large
collision systems. 
Consequently, the AMPT model has been applied to the study of various
observables in nuclear collisions such as particle yields, particle 
correlations and anisotropic flows, vorticity and polarization. 

Because the transport model approach can address non-equilibrium
dynamics, it provides a complementary framework to hydrodynamical
models for large systems at high energies, and more importantly it is
well suited to study the transition from the dilute limit to the
hydrodynamic limit. 
Therefore, it will be worthwhile to further develop a multi-phase
transport as a dynamical model for relativistic nuclear collisions. 

There are multiple areas that should be addressed in the future. 
Regarding the initial condition, at low enough energies the pure
hadron cascade version should be applicable while at high enough
energies the Eikonal formalism should be valid. 
It would be desirable to have a unified physics formulation that
self-consistently changes from one regime to the other as 
the colliding energy increases. 
In addition, for high enough energies and/or large enough collision
systems the QGP is expected to be formed, and consequently the
string melting version of the AMPT model should be
applicable instead the string-dominated default version.  
The AMPT model should be improved to dynamically determine whether the
QGP should be formed in the initial state; it would then
self-consistently change from a string-dominated initial condition to
a parton-dominated one when the initial energy density is high
enough. Another deficiency in the  initial condition of the 
string melting AMPT model is the lack of gluons in the parton phase,
and the color-glass-condensate approach would be ideal for including
initial gluons once the approach can be generalized to address the
quark degrees of freedom such as the nonzero net-baryon number.  
Regarding the parton phase, the parton cascade should be generalized
to perform transport in the presence of an electromagnetic field to
enable studies of the electromagnetic field and related observables. 
Another area of development concerns the study of
high net-baryon density physics and the QCD critical point. 
The AMPT model could be coupled to or improved with effective theories
such as the functional renormalization group method or the
Nambu-Jona-Lasinio model to treat parton interactions
self-consistently including the effective equation of state and 
effects from the critical point. 
Regarding the hadronization process, a dynamical parton recombination
criterion, e.g., by using the local parton energy density as the
recombination criterion  instead of starting hadronization at the
parton kinetic freezeout, should be developed.  Also, additional
mechanisms such as independent fragmentation should be included to
treat partons that do not find suitable coalescence partners within 
the local phase space; this would enable the AMPT model to be
applicable to studies of high $\pt$ physics once the radiative energy
loss of high $\pt$ partons is considered in the parton phase.  
Regarding the hadron cascade, it can benefit from the inclusion of
more resonances for more realistic thermodynamic properties and
chemical equilibration of the hadron matter, and modern models such as
the SMASH model could be a good choice as the new hadron
cascade component. We expect that the AMPT model in the near future,
even if only improved in a few focused areas, will enable us to
address some key questions in heavy ion physics and also serve as a
more reliable open source transport model for the community for
various studies of nuclear collisions.

\end{nolinenumbers}

\end{document}